\documentclass[reprint,longbibliography,
 amsmath,amssymb,
 prd,superscriptaddress]{revtex4-2}
\usepackage{graphicx}
\usepackage{dcolumn}
\usepackage{bm}
\usepackage[colorlinks=true, citecolor=blue, linkcolor=blue, urlcolor=blue]{hyperref}
\usepackage{xcolor}
\usepackage{tabularx}
\usepackage{amsmath, amssymb, amsfonts}
\usepackage{mathtools}
\usepackage{subfigure}
\usepackage{mathrsfs}
\usepackage{sidecap}
\usepackage{verbatim}

\newcommand{\colb}[1]{\textcolor{blue}{#1}}
\definecolor{purple1}{rgb}{128,0,128}

\usepackage{bigints}
\newcommand{\bea}{\begin{eqnarray}}
\newcommand{\ea}{\end{eqnarray}}

\definecolor{darkpastelgreen}{rgb}{0.01, 0.75, 0.24}
\begin{document}
\title{\bm{Quantum Tsunamis in a Bose-Einstein Condensate: An Analogue of Nonlinear
Gravitational Wave Memory}}
\title{Emergence of small scale structure of effective spacetime through propagation of dispersive shocks wave in Bose-Einstein Condensates} 
\title{Quantum pressure: The cosmic censor of Bose-Einstein 
condensate shock waves}
\title{The cosmic censor of shock-wave 
singularities in ultracold atomic 
condensates}
\title{Dispersive censor of 
shock-wave-singular acoustic spacetimes}
\title{Dispersive censor of acoustic spacetimes with a shock-wave singularity} 
\author{Uwe R. Fischer} 
\author{Satadal Datta}
\affiliation{%
Seoul National University, Department of Physics and Astronomy, Center for Theoretical Physics, Seoul 08826, Korea}
\date{\today}

\begin{abstract}
A dispersionless shock wave in a fluid without friction develops an acoustic spacetime singularity which is naked (not hidden by a horizon). We show that this naked nondispersive shock-wave singularity is prohibited to form in a Bose-Einstein condensate, due to the microscopic structure of the underlying ${\rm a}\!{\rm e}$ther and the resulting effective trans-Planckian dispersion. Approaching the instant of shock $t_{\rm shock}$,  rapid spatial oscillations of density and velocity develop around the shock location, which begin to emerge already slightly before $t_{\rm shock}$, 
due to the quantum pressure in the condensate.   
These oscillations render the acoustic spacetime structure completely regular,  and therefore lead to a 
removal (censoring) of the  spacetime singularity. Thus, distinct from the cosmic censorship hypothesis of Penrose formulated within Einsteinian gravity, the quantum pressure in Bose-Einstein condensates censors (prohibits) the formation of a naked shock-wave singularity, instead of hiding it behind a horizon. 
\end{abstract}

\maketitle
\section{Introduction}
In Einsteinian gravity, singularities are ubiquitous \cite{Penrose65PRL,Penrose,HawkingPredict}. 
However, the physical spacetime nature of these singularities is still under debate. 
The singularity theorems by Stephen Hawking and Roger Penrose state that if there either exists a trapped surface due to gravitational collapse or 
the Universe is assumed to be spatially closed, spacetime singularities are formed with the following conditions being satisfied: We have Einstein gravity at zero or negative cosmological constant, the weak energy condition is maintained, closed timelike curves are absent, and 
every timelike or null geodesic enters a region where the curvature is not specially alined with the geodesic 
\cite{doi:10.1098/rspa.1966.0221,cs2, 10.2307/2415769, hawking1970singularities}. {As these theorems guarantee that if there exists a trapped surface in spacetime, a singularity must form, one may ask the question if the reverse holds true,
and whether a singularity may form without a horizon enclosing it (naked singularity). The cosmic censorship hypothesis (CCH), then, 
in its weak form, states that generic gravitational collapse, starting from a nonsingular initial state, can not create a naked singularity in spacetime \cite{Penrose,Penrose99,Wald99}}. 

{However,  explicit counterexamples to the CCH, 
{for physically viable processes, have been found cf.,  e.g.,} \cite{Christodoulou1984,Roberts,PhysRevD.59.064013,PhysRevLett.99.181301,PhysRevD.79.101502}. 
On the other hand, mechanisms arguing that naked singularities are indeed hidden 
were developed, among which 
{\em backreaction} is a prominent example 
\cite{Hod,CASALS2016244,1974AnPhy..82..548W, PhysRevD.96.104014}.
It is thus fair to say that the CCH is still widely debated, as regards the possible mechanisms for either violating or preserving it, 
and whether these mechanisms are of quantum or classical origin, also cf.~Ref.~\cite{Hod2008}.   
This is largely due to the fact that there is no applicable quantum theory of gravity, in particular  
complete in the ultraviolet, with which to ascertain whether a given argument for 
(or against) the CCH is true.}

The seminal paper of Unruh \cite{unruh} triggered, especially recently, with a substantial improvement
of experimental capabilities, on a broad front a field which was coined analogue gravity \cite{BLV}. 
Its essence is that it models the propagation of classical and quantum fields 
on curved spacetime backgrounds, exploring various phenomena inaccessible at present 
in the realm of gravity proper, see, e.g., Refs.~\cite{BLV_normal,RalfBill,Weinfurtner,Euve,EuveII,Marino,Nguyen,polariton,Richartz,PhysRevD.98.064049,Andrea,Ted,Celi,Basak,Torres,Prain,
braidotti2022measurement,PhysRevD.91.124018}. A particularly promising arena are Bose-Einstein condensates
(BECs) due to the atomic precision control and accurate correlation function resolution they offer
\cite{PhysRevLett85.4643,barcelo2001analogue,Carusotto_2008,
Macher,Lahav,Steinhauer16,Munoz,PhysRevLett118.130404,
Eckel,Eckel2,Banik,Schuetzhold,BLV2003PRA,CPP,Robertson,
Gooding,Finazzi,Tian,Fedichev,Reznik,hartley2018analogue,PhysRevD.105.022003}. 

{Acoustic black holes (``dumb'' holes \cite{Unruh1994}) or cosmological horizons are thus well established and experimentally realized within the analogue gravity realm. 
On the other hand, distinct from Einstein gravity, where singularities are ubiquitous, 
singularities in quantum fluids, and with particular regard to their acoustic spacetime properties,  
have not been much studied yet, to the best of our knowledge.
It is important here to pause, and to clearly state at the outset the most important differences of analogue gravity and Einstein gravity: 
In analogue gravity, the acoustic spacetime metric is governed by nonlinear fluid dynamics 
and not by a solution of the Einstein equations. In Einstein gravity, black holes (and, as a result, also  singularities in spacetime due to the theorems by Hawking and Penrose) are formed from gravitational collapse of matter. In fluids, it is 
the transition of subsonic to supersonic flow which creates an effective dumb hole horizon for linear sound in the medium. Distinct from Einstein gravity, this analogue gravitational field, providing a 
background effective spacetime for linear perturbations on top of it, is governed by a velocity scalar \cite{Datta_2022}, in a comparable way to a 
nonlinear self-interacting scalar field theory of gravity \cite{Novello_2013}. In the present work, we establish a highly nonlinear process creating a naked singularity in the acoustic spacetime metric, 
physically represented by a  shock wave in a BEC without dispersion included (that is in the so-called Thomas-Fermi (TF) limit). 
For this nondispersive shock, the nonlinearity causes a stepwise discontinuity in the acoustic metric components, and as a result a naked timelike Ricci curvature singularity of the effective spacetime emerges. }

{In the real quantum fluid,  dispersive effects can however not be neglected, due to the quantum pressure, which occurs because of the stiffness of the condensate order parameter (scalar field) against spatial variations of its modulus.}
We reveal as a result a dispersive censorship of the spacetime singularity,  
when a nondispersive shock-wave \cite{Landau1987Fluid} would develop a singularity of the effective spacetime at its front. {Due to the dynamical differences of Einsteinian and analogue gravity based on fluid-dynamical motion, here the singularity is censored (prohibited to form), instead of being dressed by a spacetime horizon.} 
Our aim in the present investigation is thus to provide a realistic scenario, which can be experimentally implemented  in a BEC, wherein the quantum pressure censors (prohibits) the formation of a singularity in an acoustic spacetime metric. 
We therefore demonstrate that the CCH, which asserts that the naked singularity is hidden behind a horizon, is in general not necessary, provided one admits alternative theories of gravity.

\section{Fluid dynamics of dilute Bose-Einstein condensates} 
\subsection{Fluid perturbations}
Dilute BECs 
represent inviscid, barotropic, and irrotational fluids, 
where, importantly, the quantum pressure term is added to the Euler equation.  Setting the atomic mass $m=1$, we have to solve the following set 
\cite{RevModPhys.71.463}:
\begingroup
\addtolength{\jot}{0.5em}
\begin{eqnarray}
& \partial_t\rho+\nabla\cdot(\rho{\bm v})=0, \label{continuity}\\
&\partial_t{\bm v}+{\bm v}\cdot\nabla {\bm v}=-\frac{\nabla p}{\rho}+\frac{\hbar^2}{2}\nabla\left(\frac{\nabla^2\sqrt{\rho}}{\sqrt{\rho}}\right)-\nabla V_{\rm ext},\label{euler}\\
& p=p(\rho)=\frac{1}{2}g\rho^2,\label{barotropic}\\
& \nabla\times{\bm v}=0\quad \Rightarrow\quad {\bm v}=\nabla\Phi\label{rotationality}.
\end{eqnarray}
\endgroup
These equations are the only field equations occurring in our problem for condensate density $\rho({\bm r},t)$ and condensate velocity ${\bm v}({\bm r},t)$, 
and the spacetime metric for sound is then a {\em derived} and not fundamental (also see below). 
{In the above relation \eqref{rotationality},  $\Phi$ is a velocity potential due to the irrotationality of the superfluid
(excluding quantized vortex lines). The scalar potential $V_{\rm ext}$ is employed by the cold quantum gas experimentalist to create certain classes of 
effective spacetimes (see for an overview \cite{BLV}), while the condensate pressure $p$ arises 
from the two-body repulsive contact interaction between atoms, where the coefficient $g$ is proportional to the 
$s$-wave scattering length in the dilute gas \cite{RevModPhys.71.463}. Finally, the term $\frac{\hbar^2}{2}\nabla\left(\frac{\nabla^2\sqrt{\rho}}{\sqrt{\rho}}\right)$ in the Euler equation \eqref{euler}, 
is the so-called quantum pressure term \cite{RevModPhys.71.463}}. 
From the barotropic equation of state \eqref{barotropic}, the sound speed   $c_s=\sqrt{\frac{dp}{d\rho}}=\sqrt{g\rho}$; 
stability 
implies that 
$g>0$.
We linearize the fluid equations over the background of a dispersive shock wave in a BEC \cite{Damski2004}. The quantum pressure term is negligible until the shock is closely approached.
 Due to the quantum pressure term, the discontinuity in the flow, which were expected to be present in the nondispersive post-shock phase \cite{Landau1987Fluid}, is regularized. One observes instead an oscillation pattern in the density profile upon approaching the shock  (Fig.~\ref{fig4} in  {Appendix \ref{numerics}}). 
To {\it physically} distinguish classical sound wave from the background, one works with {a linear perturbation with different space and time scale than the background flow}, as discussed in the literature for linear sound propagation over background \cite{BLV}, and for nonlinear sound as well \cite{Datta_2022}. We denote background quantities with subscript ${(0)}$ and the linear perturbations with subscript $(1)$. We write ${\bm v}={\bm v}_{(0)}+\nabla\Phi_{(1)}$ by following the conventions of Ref.~\cite{Datta_2022}. {For example, with a dispersive nonlinear wave as the background, initially, 
when $t$ is much less than the shock time $t_{\rm shock}$, the wave is linear and nondispersive. For $t \ll t_{\rm shock}$, such a linear wave satisfies the massless Klein-Gordon (KG) field equation over the analogue Minkowski spacetime of a uniform static medium as background. We call this the {\em initial} background,  
and denote it with subscript $0$. {According to the Riemann wave equation for travelling one-dimensional
(1D) waves, see Eq.~\eqref{RW} below, the intrinsic nonlinearity of the fluid-dynamical equations 
becomes significant in the course of time as the wave approaches the shock \cite{Landau1987Fluid}. The KG analogy then does not hold anymore. In Ref.~\cite{Datta_2022}, we have described the  classical backreaction of the nonlinear perturbation onto the accoustic metric, and defined a {\it new} background by absorbing these nonlinear perturbations into it.} 
Here, we go near and beyond the shock time, with now in addition the quantum pressure, which originates from the spatial stiffness of the macroscopic BEC wavefunction against deformations, becoming significant.  
Linearizing 
\eqref{continuity} gives  
\begin{equation}\label{lcon}
\frac{\partial \rho_{(1)}}{\partial t}+\nabla\cdot(\rho_{(0)}\nabla\Phi_{(1)}+\rho_{(1)}{\bm v}_{(0)})=0.
\end{equation}
The linearized Euler equation follows from the Eq.~\eqref{euler}:
\begin{multline}\label{lqe}
\dot{\Phi}_{(1)}+\frac{c_{s(0)}^2}{\rho_{(0)}}\rho_{(1)}+{\bm v}_{(0)}\cdot\nabla\Phi_{(1)} \\
+\frac{\hbar ^2\rho_{(1)}}{4\rho_{(0)}^2}\left(\nabla^2\rho_{(0)}-\frac{1}{\rho_{(0)}}(\nabla\rho_{(0)})^2\right)\\
+\frac{\hbar ^2}{4\rho_{(0)}}\left(\frac{1}{\rho_{(0)}}\nabla\rho_{(0)}\cdot\nabla\rho_{(1)}-\nabla^2\rho_{(1)}\right)=0.
\end{multline}
Incorporating only the gradient  terms from the background, thus 
neglecting  $\nabla\rho_{(1)}$, and $\nabla ^2\rho_{(1)}$, we get 
\begin{equation}\label{ro1a}
\rho_{(1)} \frac{(1+ \hbar^2 \alpha) c_{s(0)}^2}{\rho_{(0)}}=-\dot{\Phi}_{(1)}-{\bm v}_{(0)}\cdot\nabla\Phi_{(1)}. 
\end{equation}
Here, we introduced a parameter $\alpha$ via  
\begin{equation}\label{alpha}
\alpha =\frac{1}{4c_{s(0)}^2}\nabla\cdot\left(\frac{\nabla\rho_{(0)}}{\rho_{(0)}}\right). 
\end{equation}
We can then define a new length scale $l=l(x,t)$ via $l_{}^{-2}\coloneqq \hbar^2 |\alpha|/\xi^2$ 
which characterizes the background spatial variation, and where 
the spatiotemporally local healing length is given by $\xi(x,t)= \xi(\rho_{(0)})=\frac{\hbar}{\sqrt{g\rho_{(0)}}}$.

The competition of the  ``microscopic"  structure dictated by $\xi$ and the ``background" scale $l_{}$ 
is expressed by $\alpha(x,t)$
which thus appears 
in the metric $q_{\mu\nu}$ in Eq.~\eqref{metricq} below.

\subsection{Spacetime metric in the dispersive fluid}

Now, we substitute $\rho_{(1)}$ from Eq.~\eqref{ro1a} into Eq.~\eqref{lcon}, 
dropping the terms in the last closed bracket of Eq.~\eqref{lqe}. 
This is the limit where the linear perturbation of all physical quantities such as $\rho_{(1)},~p_{(1)}$ can be written in terms of partial derivatives in $\Phi_{(1)}$, and the full solution can be obtained when 
$\Phi_{(1)}$ over a known background 
has been solved for. 
{Going beyond this limit requires to solve for $\rho_{(1)}$ also, and the equation of motion for $\Phi_{(1)}$ becomes an integro-differential equation \cite{barcelo2001analogue}. 
As a result, the acoustic spacetime metric 
is not local in space and time anymore. 
Here, we restrict ourselves to small wave number $k$ excitations, i.e, perturbations with wavelength larger than the coherence length $\xi(\rho_{(0)})$. In this limit, we can construct an acoustic metric local in spacetime.} 

{Linearizing in the perturbation amplitude now 
proceeds still as conventionally carried out in the analogue gravity literature \cite{unruh,BLV}. 
The difference is found in the dispersive nature of the background. The latter is 
controlled by well-posed initial (and/or boundary) conditions by the experimentalist. Over such an externally fixed, albeit nonlinear and dispersive background, {any excitation to linear order  
is called a perturbation.} {In our particular case, the highly nonlinear and dispersive background flow is clearly distinct from the linear nondispersive perturbations which experience the effective spacetime produced from such a background medium.}}  We then compare the equation of the scalar field $\Phi_{(1)}$ to  that of a minimally coupled  massless KG field equation, and find the following effective spacetime metric in 3+1D,
\begin{equation}
\label{metricq}
q_{\mu\nu}\coloneqq
\frac{\rho_{(0)}}{c_{(0)}}\begin{bmatrix}
 -(c_{(0)}^2-v_{(0)}^{2}) & \vdots & -{\bm v}_{(0)}^T \\
\cdots&\cdots&\cdots\cdots \\
-{\bm v}_{(0)}&\vdots &\mathbb{I}_{3\times 3}
\end{bmatrix},
\end{equation}
with a modified local sound speed 
\begin{equation} 
c_{(0)}=c_{s(0)}\sqrt{1+\hbar^2\alpha}
\end{equation}
due to the dispersive nature of the background. Evidently, the $\hbar^2$ small length scale correction term is present for a general background flow. 
Note that for stability, we have to impose the lower bound $\alpha >-1/\hbar^2$. 

The $q_{\mu\nu}$ are   
no longer simple algebraic functions of background density and velocity, and interpolate between
the fully nonlinear metric without dispersion $\mathfrak{g}_{\mu\nu}$ introduced in \cite{Datta_2022} 
and the {linear perturbations 
metric without dispersion}  $g_{\mu\nu}$. {See Table \ref{backgroundtable} for an overview of the various concepts and the classification of spacetime metrics in the presence of nonlinearity and/or dispersion due to quantum pressure.} {We note that the effective spacetime metric for linear perturbations 
of wavelength larger than the healing length, $q_{\mu\nu}$, does not represent a so-called rainbow spacetime \cite{Visser:2007nx,Weinfurtner_2009}. Distinct from such a rainbow spacetime,  
the metric $q_{\mu\nu}$ does not depend on the wave vector $k$ of the excitations.}

{\begin{table}[b] 
\begin{center}
\vspace*{1em}
\begin{tabular}
{|c|c|c|}
\hline
\,Background ${\rm (i)}$\,& \,${\rm{\mathfrak{Background}\,\,\mathfrak{(ii)}}}$\,& \,Background ${\rm (iii)}$\,
\\\hline
$\rho_0, {\bm{v}}_{0}$ & $\rho_{(0)}, {\bm{v}}_{(0)} $  
&  $\rho_{(0)} $, ${\bm{v}}_{(0)}$ \\
$l_{}\gg\xi$ & $l_{}\gg\xi$  & $l_{}\sim\xi$ \\
\hline  
 $g_{\mu\nu}$ &   ${\mathfrak g}_{\mu\nu}$  & $q_{\mu\nu}$ \\
 \hline
\end{tabular}
\caption{{Defining background flows from nonlinearity and dispersion 
and their associated metrics, where $l_{}$ is the length scale defined below 
\eqref{alpha}. 
For Background (i), $\rho_{0}, \bm{v}_{0}$ represent a solution of the nondispersive fluid equations without quantum pressure, and are initially chosen as the background before the shock develops, with perturbations treated to linear order. This initial Background (i) corresponds to the conventional analogue gravity metric and 
may or may not derive from nonlinear fluid equations; for example a uniform static medium  
does not represent a nonlinear background. 
For the $\rm{\mathfrak{Background}\,\,\mathfrak{(ii)}}$, $\rho_{(0)}, \bm{v}_{(0)}$, are found from the fully nonlinear, coupled  fluid equations for both background and perturbations, {however without quantum pressure included.} 
cf.~Ref.~\cite{Datta_2022}. 
Finally, for the Background (iii), $\rho_{(0)}, \bm{v}_{(0)}$ are found {from the nonlinear fluid equations
applied to the background motion alone, but now with quantum pressure included. } 
}}
\label{backgroundtable}
\end{center}
\end{table}}

\section{Dispersive shock waves}  
We consider the propagation of a wave, initially created as a Gaussian distribution, in the condensate.  
We consider a realistic situation, with the effect of quantum pressure included, i.e., a highly nonlinear dispersive wave \cite{Damski2004}. 
The acoustic metric of such nonlinear dispersive pulse wave in our quasi-1D BEC set up, is given by the Eq.~\eqref{metricq} with ${\bm v}_{(0)}$ having only one component along $x$ axis, $v_{(0)}(x,t)$.

We choose the initial wave profile \cite{Damski2004} as the Gaussian 
\begin{equation}
\begin{aligned}
& \rho_{(0)}(x, t=0)=\rho_\infty\left(1+2\eta \exp\left[-\frac{x^2}{2\sigma ^2}\right]\right), \\
& v_{(0)}(x, t=0)=0, \label{roin} 
\end{aligned}
\end{equation} 
where $\sigma \gg \xi(\rho_{(0)})$. 
Here, at the center of our quasi-1D BEC set up, we produce a source of gravitational wave (GW) with density being almost uniform towards the boundary, mimicking asymptotically flat effective spacetime with a GW source. {This longitudinal GW  is different from its counterpart in Einstein gravity, in that the spacetime lacks general covariance, and the GW can not be represented in its usual transverse and traceless form, cf.~the discussion in \cite{PhysRevD.105.022003}. }

The Thomas-Fermi profile in Eq.~\eqref{roin}
(neglecting the impact of quantum pressure on the initial state) 
can be created by focusing a laser detuned from atomic resonance onto 
 the center of the one dimensional condensate, with a size $\gg\sigma$ \cite{Damski2004}.  Switching off the laser creates a nonlinear dispersive propagating wave with {high frequency} oscillations when the shock occurs,
 as previously described in \cite{Damski2004}, see for a detailed description Appendix \ref{numerics}.  
{Shock waves in quasi-1D BECs have been experimentally observed \cite{meppelink2009observation}, and also in nonlinear photon fluids  \cite{marino2016emergent}. 
In particular, Ref.~\cite{meppelink2009observation} captures density modulations which may be compared to the 
high-frequency post-shock oscillations predicted by Damski \cite{Damski2004}.} 

{We numerically solve the fluid equations (in a box potential with $\nabla V_{\rm ext}=0$), that is 
Eqs.~\eqref{continuity} to \eqref{rotationality}, employing a 4th order Runge-Kutta method to perform the time integration, and expanding the spatial derivatives within a central difference method scheme up to the same 4th order accuracy \cite{shen2016introduction}. } 
\begin{figure}[t]
\includegraphics[scale=0.275]{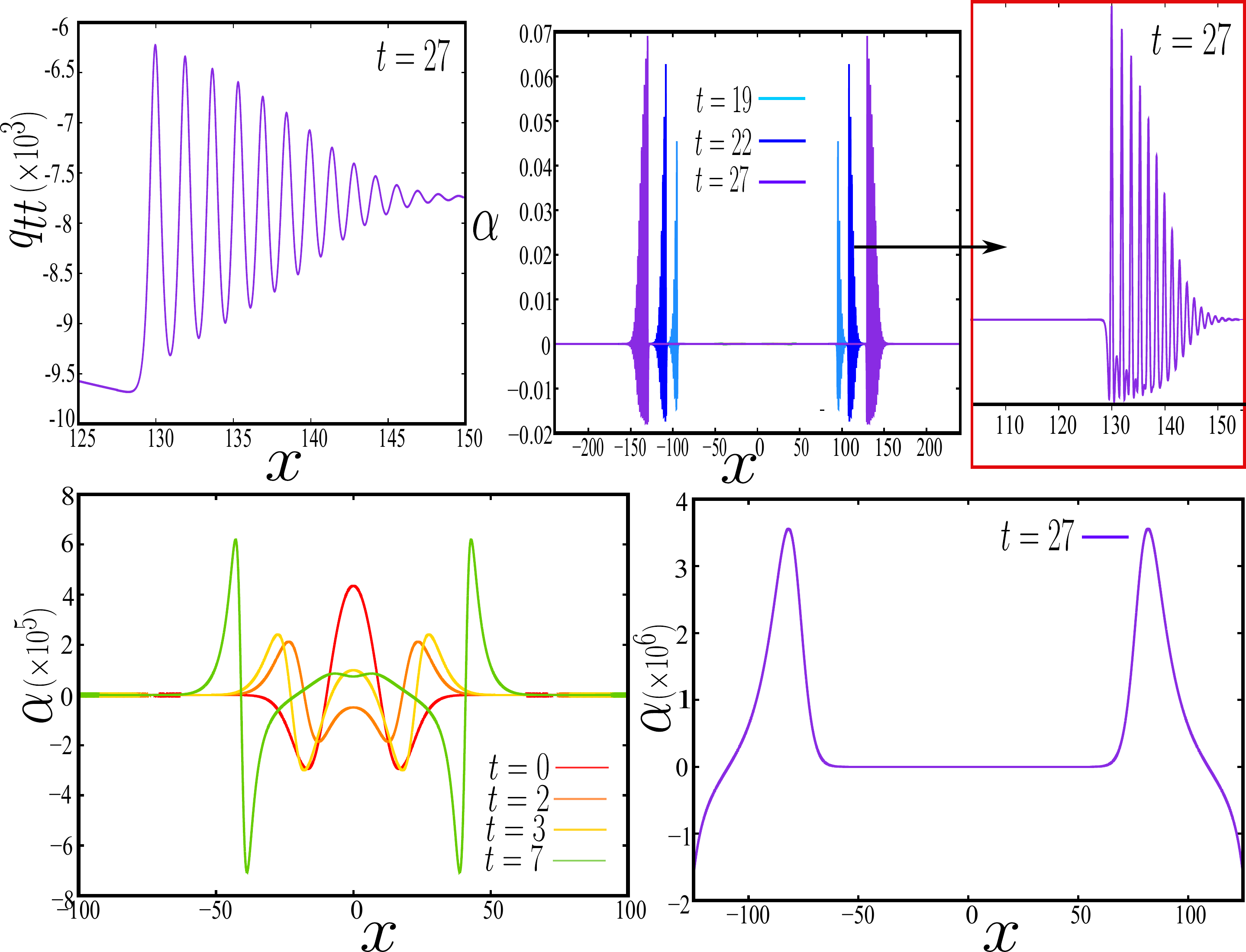}\

  \centering
  
   \caption{{Behavior of $q_{\mu\nu}$ for the dispersive shock wave as background, found by numerically solving Eqs.~\eqref{continuity}-\eqref{rotationality}.}
    {We use here units in terms of a {length} $L$, chosen appropriately for the purpose of our 
    numerical calculation.  {For example, in the experiment \cite{meppelink2009observation} the size of the condensate is roughly $500\,\mu m$, whereas the full x-axis range of our numerical simulation is 500. Therefore $L$ would be approximately $1\,\mu m$ with the parameters of Ref.~\cite{meppelink2009observation}.} 
    Time $t$ is then measured in units of $L^2$, when setting $\hbar=1$. In these units,} 
the  parameters we choose are $g=7500$, $\rho_\infty=0.002$, $\eta=0.2$, $\sigma=8.838$.
{We then obtain $t_{\rm shock}\simeq 13.43$, using the method described in Ref.~\cite{Damski2004}.}  Only after and slightly before the instant of shock, the dispersive nature of the background flow  becomes important in the oscillatory region, and hence in $q_{\mu\nu}$. Top left (at $t=27$): 
We plot $q_{tt}$ in the post-shock phase; the behavior of other metric components is similar. 
 (Bottom row):  In the post-shock phase, the amplitude of the parameter $\alpha$ defined in Eq.~\eqref{alpha} in 
 the nonoscillatory region is essentially negligible compared to its amplitude in the oscillatory region;  it however increases rapidly as $t$ approaches $t_{\rm shock}$.}
\label{fig1}
 \vspace*{-1.5em}
\end{figure}
{We now consider only nonlinearity taken into account for the fluid motion, i.e., Eq.~\eqref{continuity}-\eqref{rotationality} without quantum pressure, and with the initial profile of Eq.~\eqref{roin}. 
After a certain time, the initial Gaussian density wave profile separates completely into two identical smaller pieces (while respecting mass conservation), and moving in opposite directions. 
The right-moving travelling wave in the polytropic gas with pressure 
$p_{(0)}\propto \rho_{(0)}^\gamma$ (for BECs $\gamma=2$) can be described in terms of single variable $v_{(0)}(x,t)$ by the Riemann wave equation \cite{Riemann1860}:}
\begin{eqnarray}
\frac{\partial v_{(0)}}{\partial t}+\left[c_{s0}+ \left(\frac{\gamma +1}{2}\right)v_{(0)}\right]\frac{\partial v_{(0)}}{\partial x}=0,\label{RW}\\
\rho_{(0)}=\rho_{0}\left[1+\left(\frac{\gamma -1}{2}\right)\frac{v_{(0)}}{c_{s0}}\right]^{\frac{2}{\gamma-1}}. \label{vro}
\end{eqnarray}
{The second identity directly relating density to flow speed perturbations is valid for a {\em simple wave}  \cite{Landau1987Fluid}.}  
The left-moving travelling wave comes with a $`-$' sign in front of $c_{s0}$ in the above equations;  
$\rho_{(0)}=\rho_\infty$ for $v_{(0)}=0$, $\rho_\infty \simeq \rho_{0}$ of Eq.~\eqref{roin} since $\sigma \ll $ size of the condensate. 
This first-order quasi linear partial differential equation leads to multivalued valued solution by the method of characteristics \cite{PDES}.

By obeying momentum and mass conservation across the discontinuity, one is led to the equal area rule { $\oint (x-x_s)dv_{(0)}=0$ where $x_s$ is the shock location (location of discontinuity)}, to avoid such a multivalued solution from the  shock time ($=t_{\rm shock}$) onward \cite{Landau1987Fluid}. We discuss this issue further in Appendix \ref{nondispersiveshock}. In the presence of quantum pressure, the solution (density, velocity etc) becomes oscillatory around the discontinuity, in comparison in the Fig.~\ref{fig2}. Therefore, the solution becomes a {\it well behaved} function of $x$ and $t$ \cite{Damski2004}, see 
{Appendix \ref{numerics}}. Numerical solution of Eq.~\eqref{continuity} together with Eq.~\eqref{rotationality} produces $q_{tt}$ in Fig.~\ref{fig1}. As expected, $\alpha$ is practically zero in the nonoscillatory region. 
The $\alpha$-correction term in the metric $q_{\mu\nu}$, which is usually hidden in a slowly varying background, is amplified in a region where quantum pressure is important: It is a significant contribution relative to the other forces in the Euler-type evolution of momentum \eqref{euler} in the oscillatory region (cf.~Fig.~\ref{fig5} in the {Appendix \ref{numerics}}).  
Remarkably, the oscillations in the solution starts just slightly before the shock time $t_{\rm shock}$ (see Fig.~\ref{fig7} 
 {Appendix \ref{numerics}}), whereas $ t_{\rm shock}$ is computed in the zero quantum pressure limit. {Therefore,} $t_{\rm shock}$ maintains its importance as a time scale even  with quantum pressure, signifying the time of initiation of oscillation. 
A linear travelling 1D wave can not stay linear forever, after a certain time nonlinearity makes the $v_{(0)}$ profile steeper,
 with negative $\frac{\partial v_{(0)}(x,t)}{\partial x}$. This renders, in turn, the quantum pressure significant. Thus nonlinearity invites dispersion due to quantum pressure to play a significant role, {also see the Appendices \ref{nondispersiveshock} and \ref{numerics}.}
\section{Censoring the Naked Singularity} 
\begin{figure}[t]
\centering
\includegraphics[scale=0.165]{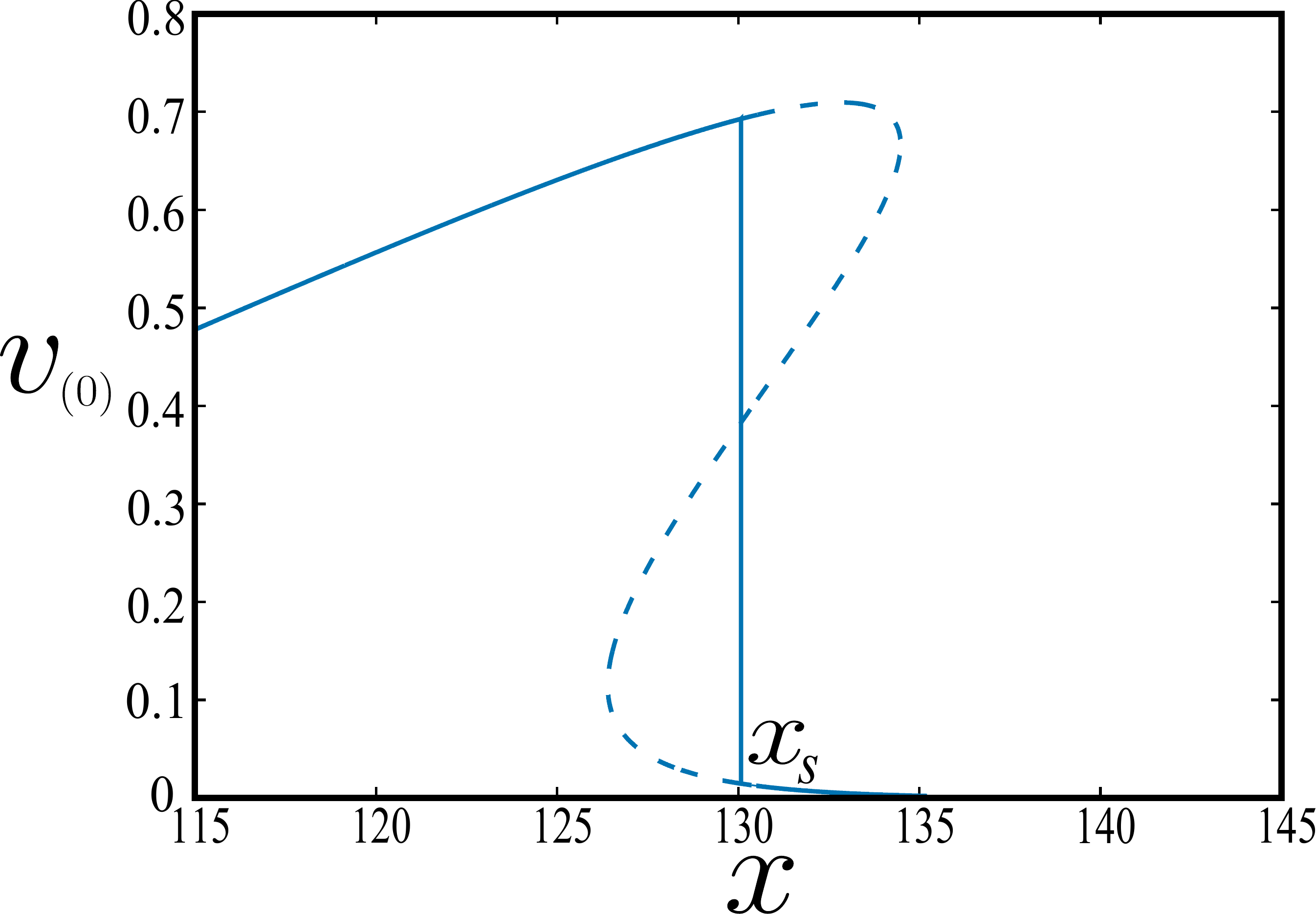}\includegraphics[scale=0.165]{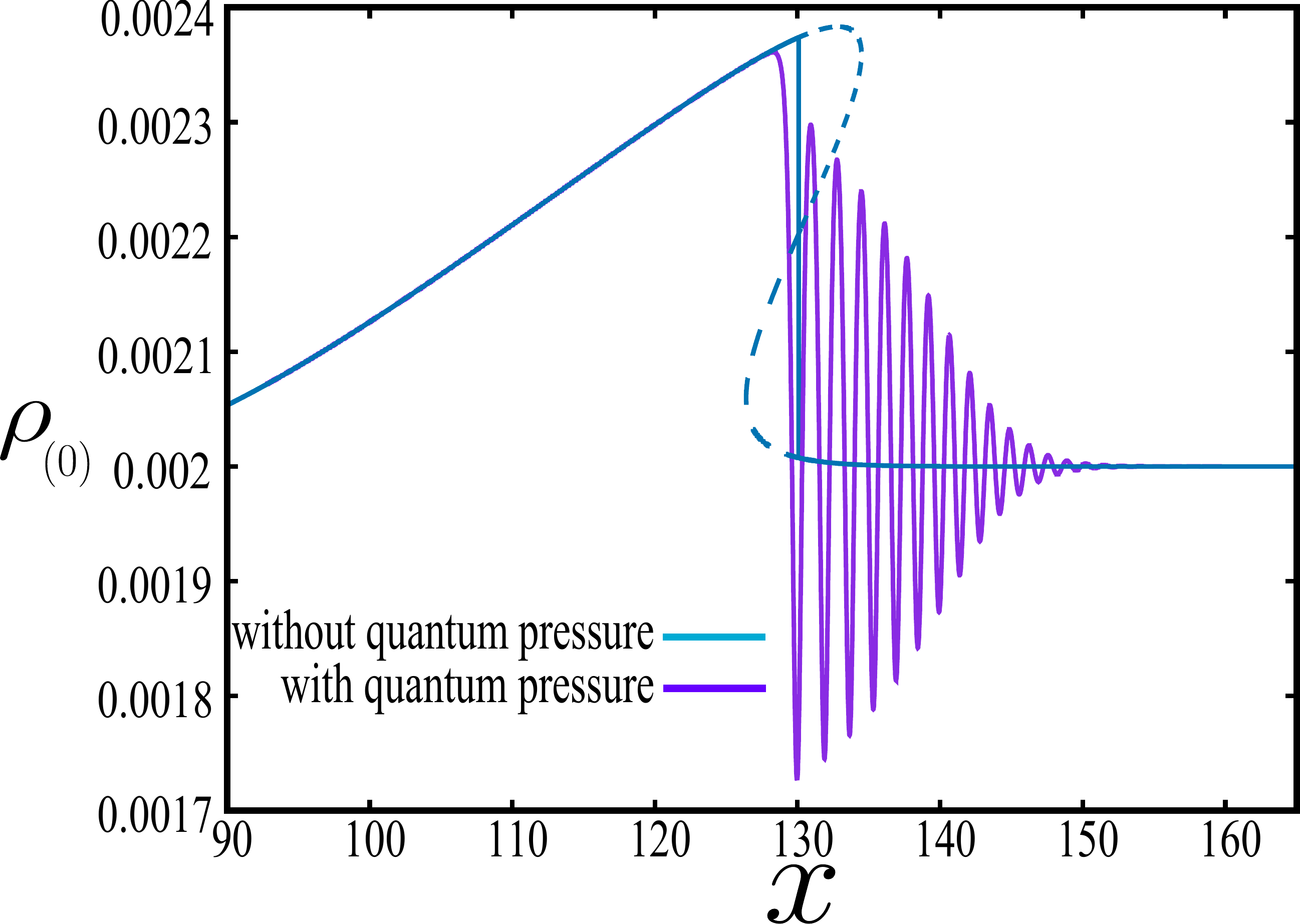}
  
  
   \caption{{(Left) Multivalued solution of postshock Riemann wave by the method of characteristics (dotted), physical solution with discontinuity (nondispersive shock)  by equal area rule \cite{Landau1987Fluid} (solid line).
  (Right) Discontinuity in the flow is avoided as the wave approaches the shock time when we take quantum pressure into account. Inclusion of quantum pressure in the equation creates oscillation and thus continuous solution of $v_{(0)}$ removes singularity in the acoustic metric, i.e., censorship of singularity. The wave profile with quantum pressure is a good match with the nondispersive nonlinear wave profile in the region except the rapidly oscillatory region. Parameters as in Fig.~\ref{fig1}.}}
\label{fig2}
\vspace*{-1.5em}
\end{figure}
We now aim to find what a discontinuity in the solution 
means for the effective spacetime.
We denote the acoustic metric for the nondispersive metric as $\mathfrak{g}_{\mu\nu}$, cf.~Table~\ref{backgroundtable}. 
We stress that,  while the metric is derived nondispersively, it is still taking the nonlinearity of the fluid into account \cite{Datta_2022}. It reads 
\begin{multline}\label{lineds}
ds^2=\mathfrak{g}_{\mu\nu}dx^\mu dx^\nu
=\frac{\rho_{(0)}}{c_{s(0)}}
\left[-(c_{s(0)}^2-v_{(0)}^2)dt^2\right.\\
\left.-2v_{(0)} dtdx+\sum_{i=1,2,3}({dx^i})^2\right] .
\end{multline}
Note that this metric 
is also {\em not} identical to the conventional analogue gravity metric $g_{\mu\nu}$, 
which assumes that the dynamics of perturbations is linear instead of nonlinear, cf.~Table~\ref{backgroundtable}
for a classification of metrics.  The linear approximation is valid only for small amplitudes and short time intervals, while 
the quantities $\rho_{(0)}$, $v_{(0)}$ in the metric are found from the solution of the nonlinear fluid equations without quantum pressure. 
This is  $\rm{\mathfrak{Background}\,\,\mathfrak{(ii)}}$ in Table \ref{backgroundtable}. For nonlinear dispersive shock wave,  $\rm{\mathfrak{Background}\,\,\mathfrak{(ii)}}$ and Background (iii) coincide very well in every region except in the oscillatory region, i.e., the region around shock location $x_s$. In the asymptotic region, i.e., near the condensate wall, 
$\rm{\mathfrak{Background}\,\,\mathfrak{(ii)}}$ and Background (iii) coincide with the Background (i) which is uniform and static, i.e., an acoustic analogue of Minkowski spacetime.

Evidently the acoustic metric is discontinuous at $x=x_s$ after  the shock has occurred. We compute the Ricci scalar, $ R$ \cite{weinberg1972gravitation} 
 {for $\mathfrak{g}_{\mu\nu}$ for the right moving travelling wave satisfying Eq.~\eqref{RW}}.
We perform the calculations in Mathematica, replacing 
$\partial_t$ by $\partial_x$ derivatives, employing the Riemann wave equation \eqref{RW}. This procedure
leads to the surprisingly simple relation 
\begin{eqnarray}\label{RicciRich}
& R=\frac{\textstyle(1+\gamma)}{\textstyle\rho_{(0)}}\frac{\textstyle\partial^2 v_{(0)}(x,t)}{\textstyle\partial x^2}. \label{Rc}
\end{eqnarray}
expressing the curvature scalar solely by the second spatial derivative of the background flow field. 
At $x=x_s$, $v_{(0)}=v_{1}$ and $\rho_{(0)}=\rho_{1}$ which are the pre-shock values of velocity and density respectively, related to each other by the Eq.~\eqref{vro}. Since in this case, the wave is propagating from left to right, at $x=x_s$, $v_{(0)}$ first has $v_1$ then it jumps to post-shock value $v_{2}~(<v_1)$, thus unrealistic multivalued $v_{(0)}$ is avoided. \colb{$\lim_{x\to x_s}v_{(0)}(x,t)$} doesn't exist, but it has a definite value which is $v_1$, and as a consequence; this discontinuity can be written mathematically in terms of a Heaviside step function, {see Appendix \ref{nondispersiveshock}}.  $\frac{\partial v_{(0)}(x,t)}{\partial x}=-\infty$ at $x=x_s$, and $\frac{\partial^2 v_{(0)}(x,t)}{\partial x^2}$ {at $x=x_s$} can be expressed as a summation of $\delta (0)$ and $\delta '(0)$ (with definite coefficients) type of infinities {(in Appendix \ref{nondispersiveshock})}; where $'$ denotes a $x$ derivative. 
We discuss the visualization of Dirac delta distributions through a delta-sequence function in Fig.~\ref{fig3} of Appendix \ref{nondispersiveshock}. 

{We plot in Fig.~\ref{figr} the Ricci scalar of the nondispersive wave as it approaches the curvature singularity in the pre-shock phase $t<t_{\rm shock}$.}
As can be seen, the expression \eqref{Rc} implies the existence of a (strong) curvature singularity at $x=x_s$, where $x_s$ is the position of discontinuity at $t\geq t_{\rm shock}$.}
\begin{figure}[t]
\includegraphics[scale=0.275]{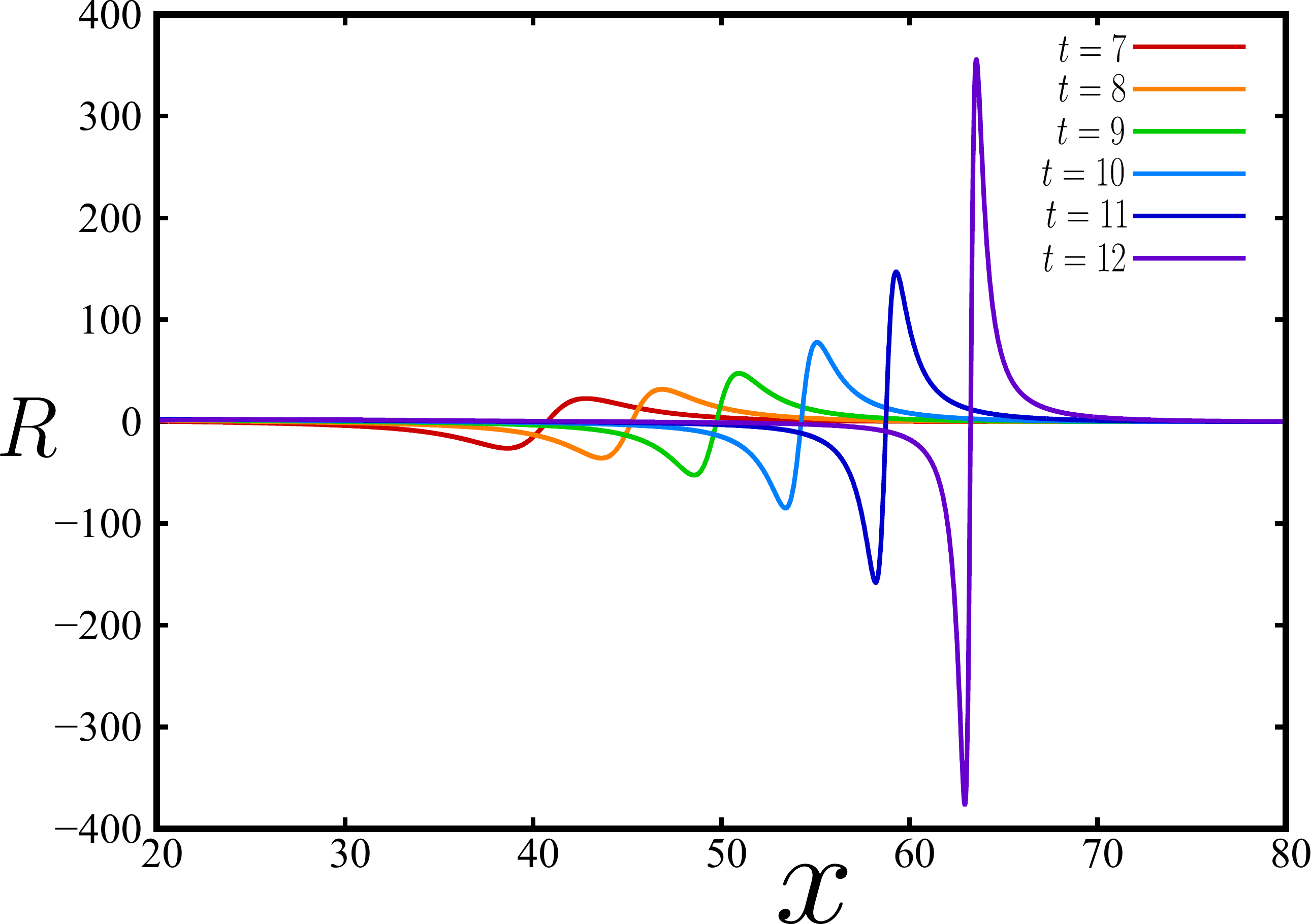}
  \centering
   \caption{The Ricci scalar of $\mathfrak{g}_{\mu\nu}$ approaching the  shock singularity of 
   the nondispersive Riemann wave.   Parameters are identical to those of Fig.~\ref{fig1}.}
\label{figr}
\vspace*{-2em}
\end{figure}
Since the velocity at any $x$ remains always very much less than the minimum value of sound speed $c_{s0}$ $(=\sqrt{g\rho_0})$,  there is no event horizon present in the acoustic metric. Since at $x=x_s$, $v_{(0)}=v_1$; sound speed $c_{s(0)}=c_{s1}=c_{s0}+\left(\frac{\gamma +1}{2}\right)v_{1}$, and the travel speed of the discontinuity is $u=c_{s0}+\left(\frac{\gamma +1}{4}\right)(v_{1}+v_{2})$, $v_{2}$ $(<v_{1})$ is the post-shock value of $v_{(0)}$ \cite{Landau1987Fluid}. Hence $c_{s1}> u$. In Eq.~\eqref{lineds}, by putting $dx =udt$, $dy=dz=0$, we find
$ds^2=\frac{\rho _1}{c_{s1}}\left(-c_{s1}^2+(u-v_1)^2\right)dt^2$, from the above discussion, we notice that $c_{s1}>|u-v_1|$. Therefore, at $x=x_s$, the discontinuity follows a timelike trajectory, representing a naked singularity. 
When we, on the other hand, solve the fluid equations with quantum pressure, the solution oscillates instead of discontinuity, we render the curvature for the metric $q_{\mu\nu}$ $\forall$ $x$ and $t$ finite, thus removing the singularity, cf.~Fig.~\ref{fig2}. However, for nondispersive waves,  
the discontinuity 
does not persist for $t\to\infty$, and 
$(v_1-v_2)$ then falls to zero 
\cite{Landau1987Fluid}.

\section{Comparison with hydraulic jump} 
To put the above discussion on dispersively censoring shock-wave spacetime singularities 
in perspective, we compare it with another example for a possible spacetime 
singularity, now in classical liquids, the so-called hydraulic jump \cite{rayleigh}. 

In general relativity, the components of the spacetime metric reflect a choice of coordinate system, and there is no preferred coordinate system. Hence constructing scalar quantities, such as the Ricci scalar, quantifying the curvature is important to distinguish genuine spacetime singularities from singularities removable by coordinate transformations. However, the acoustic metric components for analogue gravity (in the present nonrelativistic background framework) are functions of physical quantities (velocity and density of the background flow). Then, a discontinuity in the acoustic metric can also be regarded as a {\it physical} singularity, i.e., some kind of boundary between two different spacetime manifolds. Therefore, the singularity for the post-shock simple wave is not only a Ricci scalar curvature singularity at $x=x_s$, but also can be regarded as the boundary between two different manifolds with two distinct acoustic spacetime metric defined on them. 

The hydraulic jump 
possesses a {\it physical} singularity at an effective radial white hole horizon (for the circular hydraulic jump), 
as represented by a sudden increase in {fluid height} at the circular boundary 
\cite{vHe}. The white-hole horizon for the circular jump has for example been experimentally studied for 
{viscous silicon oil with low surface tension (and with therefore no capillary dispersion)} 
 \cite{gr1}. 
Distinct from the singularity for shock wave, the singularity for the hydraulic jump is neither naked nor hidden behind a horizon,  as the hydraulic jump spacetime singularity occurs exactly at the horizon \cite{vHe}.  
If the hydraulic jump 
is ``noticeably" sharp, as observed in liquid Helium \cite{He} as well as in viscous silicon oil \cite{gr1}, such a jump can indeed be considered a {\it physical} singularity. 
However, the continuum approximation in fluid dynamics is valid over after coarse-graining over a certain length scale. For example, in the case of the flow of a real gas, the fluid descriptions of physical quantities such as velocity and density  are valid on a  length scale much bigger than the mean free path of the constituent particles. Similarly, for a BEC with quantum pressure included, the number of atoms per healing length 
has to be much greater than unity for the mean-field hydrodynamical description to apply. 
Therefore, the description in terms of a spacetime singularity due to a discontinuity in the background flow holds on the length scales for which fluid dynamics is valid.

In rectangular channel flows, the jump in {fluid height} is however noticeably smooth instead of sharp \cite{gr2,PhysRevLett.124.141101}, and for narrow channel flow, the hydraulic jump is followed by a post-jump undulation, 
constituting the so-called {\em undular} hydraulic jump \cite{chr,gr3}. 
The undular hydraulic jump has been studied in viscous flows, e.g. in  
 \cite{viscousjump},  
as well as turbulent flows, e.g. in  
\cite{Steinruck_2003}. 
The dissipation due to turbulence and viscosity for the channel undular hydraulic 
jump prevents a sharp rise in {fluid height.}  

The dispersive shock wave problem that we consider here for a BEC is structurally similar to the Korteweg–De Vries  
equation, which includes nonlinearity and dispersion \cite{kam,gurevich}. In our case, the dispersion is due to quantum pressure, which modifies the acoustic metric Eq.~\eqref{metricq}, and, as a consequence, resolves the singularity in the metric. By contrast, the undular jump in channel flows involves dissipation in addition, which is complicating its analysis.

To summarize, in distinction to the (undular) hydraulic jump, in our simplified 1D shock-wave setup we have 
 no turbulence (flow speeds remaining well below the speed of sound), and no spacetime horizon.  
 We also have no dissipation for a BEC at $T=0$. 
 Finally, the dispersion we consider in a BEC, while in the shallow water limit formally similar to quartic order in wavenumber, has a different physical origin than for the hydraulic jump \cite{gr1}. 
 Finally, as far as we are aware, our study 
presents the first confirmation of a {\em spacetime} singularity by explicitly calculating the corresponding divergence of the Ricci curvature scalar. 
\bigskip

\section{Conclusion} 
We demonstrated that the quantum pressure term  leads to a regular oscillatory numerical solution for travelling waves in a quasi-1D BEC, thus prohibiting the otherwise naked singularity. 
{Analogue gravity is effectively an ${\rm a}\!{\rm e}$ther theory, for which we have shown, 
using a particular initial condition, that the occurrence of a naked singularity is forbidden.
Whether singularities in the dispersive ${\rm a}\!{\rm e}$ther  of the 
BEC arise for {\em any} given nonsingular initial condition is an open question.} 

{We have thus provided, for a BEC laboratory analogue simulating curved spacetimes,
a censor prototype operating in the trans-Planckian sector of the dispersion relation,
which is based on the microscopic 
physics of the system, and is thus naturally complete in the ultraviolet. 
To ultimately resolve the question of whether the CCH holds true, this latter property is crucial also for any proper quantum gravity.
}

\acknowledgements  
We thank B. Damski and F. Marino for helpful discussions on dispersive shock waves.  
This work has been supported by the National Research Foundation of Korea under 
Grants No.~2017R1A2A2A05001422 and No.~2020R1A2C2008103.


\appendix 
\section{Nondispersive shock waves and the curvature singularity}
\label{nondispersiveshock}
{In this Appendix, first we briefly introduce the equal area principle introduced in 
\cite{Landau1987Fluid} for nondispersive shock waves, and then we proceed to calculating the Ricci scalar curvature for such a nondispersive shock wave.}

\begin{figure}[t]
\centering
\includegraphics[scale=0.15]{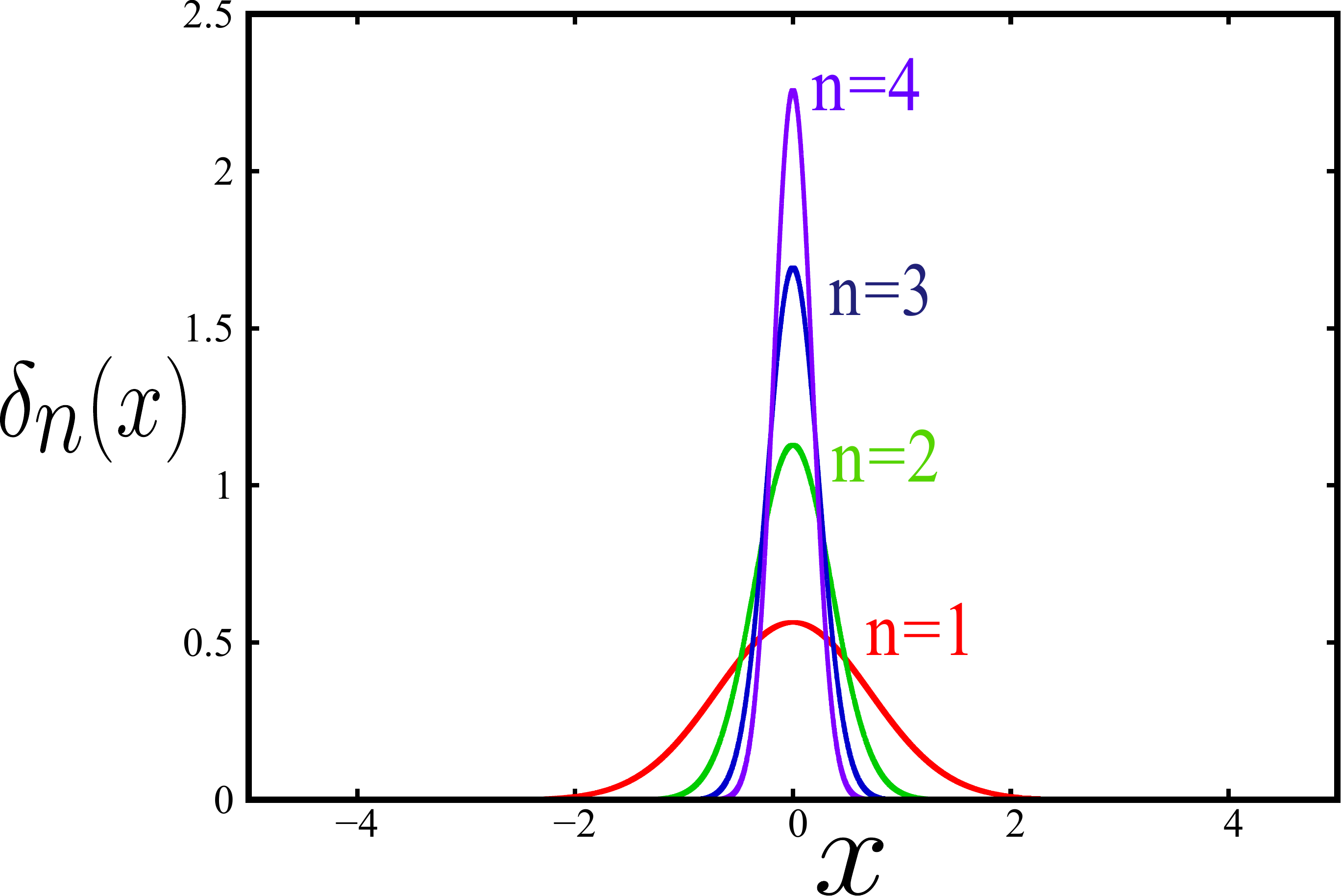}\includegraphics[scale=0.15]{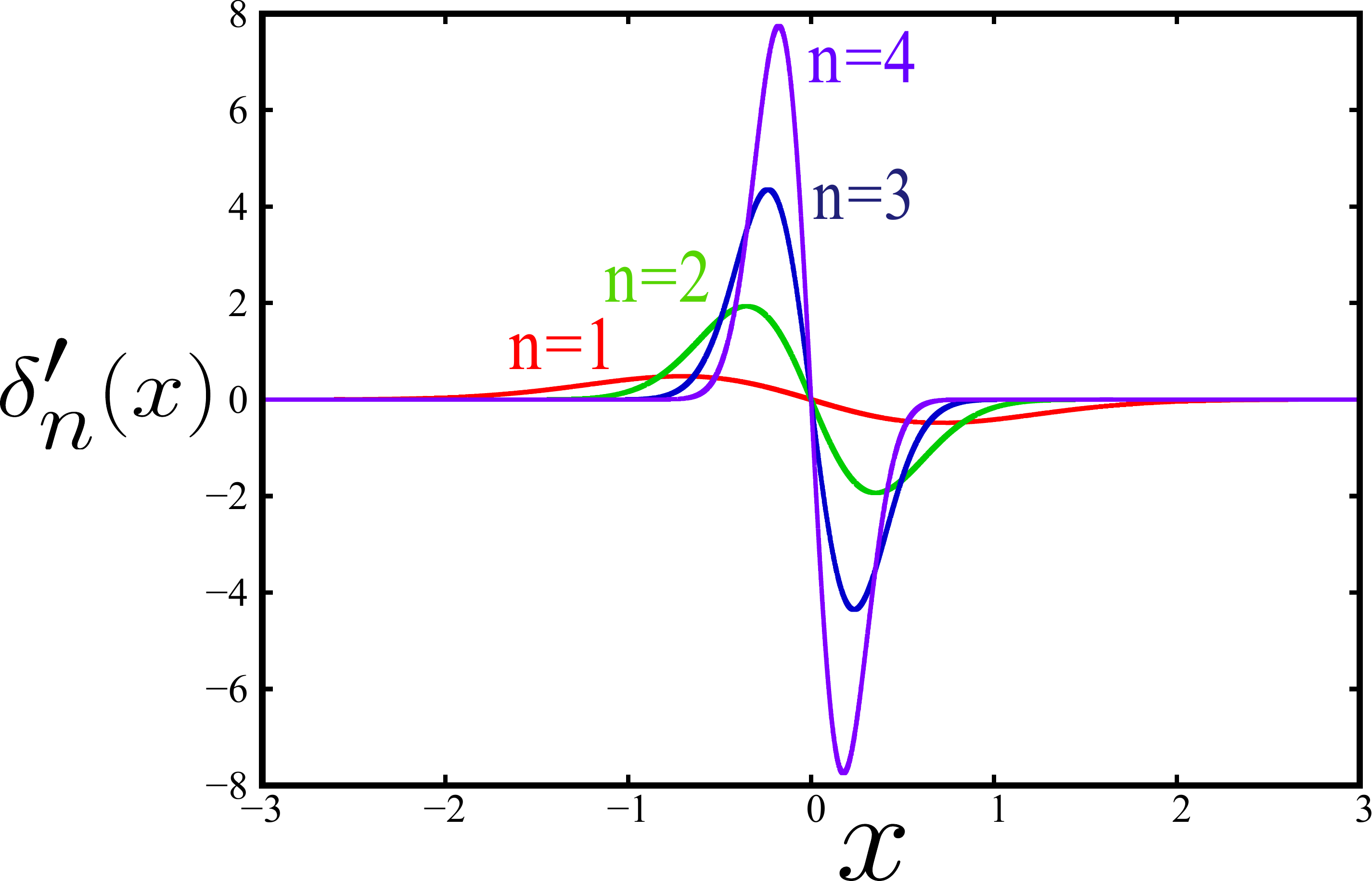}
  
   \caption{Sequences up to $n=4$ are shown to represent the delta distribution function (left) and its first derivative (right). $\displaystyle{\lim_{n \to \infty}}\int_{-\infty}^{\infty}f(x)\delta_n(x)dx=\int_{-\infty}^{\infty}f(x)\delta (x)dx$, and $\displaystyle{\lim_{n \to \infty}}\int_{-\infty}^{\infty}f(x)\delta'_n(x)dx=\int_{-\infty}^{\infty}f(x)\delta'(x)dx$ (for arbitrary $f(x)$) are used to define a relation of the Dirac delta distribution and its derivative with their respective sequence functions \cite{arfken2013mathematical}.  
}
\label{fig3}  
\end{figure}

The Riemann wave Eq.~\eqref{RW} can be solved by the analytical techniques for partial differential equations, i.e., the method of characteristics. This analytical solution \cite{PhysRevD.105.022003} gives rise to multivalued solution after a certain time, $t_{\rm shock}$. At $t=t_{\rm shock}$, $\frac{\partial v_{(0)}(x,t)}{\partial x}$ reaches infinity \cite{Landau1987Fluid}. If we follow the method of characteristics \cite{Landau1987Fluid,Datta_2022} to solve Eq.~\eqref{RW} for the case without quantum pressure to avoid multivalued solution of density and velocity after $t_{\rm shock}$, the solution has to become discontinuous. This jump in velocity (and density) approximately satisfies the equal area rule \cite{Landau1987Fluid}: 
\begin{equation}\label{equalarea}
\int_{v_1}^{v_2}(x-x_{s})dv_{(0)}=0,
\end{equation}
where $v_1$ and $v_2$ ($v_1>v_2$) are the pre-shock and post-shock values of discontinuous velocity $v_{(0)}$ across the position of discontinuity (shock) at $x=x_{s}$. As a result, $\rho_1$ and $\rho_2$ are pre-shock and post-shock values of density $\rho_{(0)}$ related to $v_1$ and $v_2$ by 
\begin{eqnarray}
\label{rhov12}
\rho_{1,2}=\rho_{0}\left[1+\left(\frac{\gamma -1}{2}\right)\frac{v_{1,2}}{c_{s0}}\right]^{\frac{2}{\gamma-1}}.
\end{eqnarray}
With this discontinuity, velocity and density profiles are not multivalued anymore, which is discussed 
in detail by the classic textbook \cite{Landau1987Fluid} . {The expression of Ricci scalar (Eq. \eqref{RicciRich}) in the nondispersive limit is proportional to the second derivative in $v_{(0)}$, here we discuss an analytical way to calculate the second derivative of $v_{(0)}$ with a discontinuity at $x=x_s$.}
This discontinuous velocity profile $v_{(0)}(x,t)$ at fixed time $t>t_{\rm  shock}$ can written in a compact approximate way,
\begin{equation}\label{fcurves}
v_{(0)}(x,t)=\left(1-\Theta (x-x_s)\right) f_1(x)+\Theta (x-x_s)f_2(x),
\end{equation}
where $\Theta$ is the Heaviside step function, defined by 
$\Theta (x-x_s)=1~~{\rm for}~ x>x_s$ and 
$\Theta (x-x_s)=0~~{\rm for}~ x\leq x_s$. 
Furthermore, $f_1(x)$, $f_2(x)$ are Newton interpolation polynomials \cite{shen2016introduction}, constructed from a finite number of points on the pre-shock curve segment and on the post-shock curve segment of $v_{(0)}(x,t)$ respectively, at a fixed time $t>t_{\rm shock}$, e.g., from the left subfigure of the Fig.~\ref{fig2}. Thus we approximately describe $v_{(0)}(x,t)$ at fixed $t>t_{\rm shock}$ by these two polynomials with finite coefficients in a compact way. Therefore, $f_1(x)$ and $f_2(x)$, for {\it a reasonably accurate} fitting, should satisfy (a) $f_1(x_s)\sim v_1>f_2(x_s)\sim v_2$, and (b) the slopes of $f_1(x)$ and $f_2(x)$, at $x=x_s$ {\it smoothly} fits into the pre-shock curve segment and post-shock curve segment, respectively.
We find
\begin{multline}
\frac{\partial v_{(0)}}{\partial x}=\left(1-\Theta (x-x_s)\right)\frac{df_1}{dx} \\
+\Theta (x-x_s)\frac{df_2}{dx}+\delta(x-x_s)\left(f_2(x)-f_1(x)\right), 
\end{multline}
where $\delta (x-x_s)$ is the Dirac delta distribution. The first two finite terms of the equation has a similar pattern to the Eq.~\eqref{fcurves} for obvious reasons. Therefore,
\begin{equation}
\frac{\partial v_{(0)}}{\partial x}|_{x=x_s}=\delta (0)(v_2-v_1)+\frac{df_1}{dx}|_{x=x_s}.
\end{equation}
Evidently, the first term on the right hand side dominates over the second term, rendering $\frac{\partial v_{(0)}}{\partial x}|_{x=x_s}$ to be $-\infty$, since $v_2<v_1$.
\begin{multline}\label{2dd}
\frac{\partial^2 v_{(0)}}{\partial x^2}=\left(1-\Theta (x-x_s)\right)\frac{d^2f_1}{dx^2}
\\
+\Theta (x-x_s)\frac{d^2f_2}{dx^2}+2\delta(x-x_s)\left(\frac{df_2}{dx}-\frac{df_1}{dx}\right)
\\
+\delta '(x-x_s)\left(f_2(x)-f_1(x)\right)
\end{multline}
Therefore, at $x=x_s$, ignoring the finite term $\frac{d^2f_1}{dx^2}|_{x=x_s}$, we write down the infinite terms as follows, 
\begin{multline}
\frac{\partial^2 v_{(0)}}{\partial x^2}|_{x=x_s} \\
=2\delta(0)\left(\frac{df_2}{dx}|_{x=x_s}-\frac{df_1}{dx}|_{x=x_s}\right)+\delta '(0)\left(v_2-v_1\right).
\end{multline}
According to Fig.~\ref{fig2}, $\frac{df_2}{dx}|_{x=x_s}$ is negative; it always stays negative in the post-shock phase, and $\frac{df_1}{dx}|_{x=x_s}$ is positive. Numerics in fact shows that, initially after $t_{\rm shock}$, $\frac{df_1}{dx}|_{x=x_s}$ is negative, but eventually it becomes positive over time. {The quantity $\frac{\partial^2 v_{(0)}}{\partial x^2}|_{x=x_s}$ above consists of two different kinds of infinity. One can represent them by $\delta$-sequence functions \cite{arfken2013mathematical}. We choose here a particular one to describe these infinities (see also Fig.~\ref{fig3})}, 
\begin{eqnarray}
& \delta _n(x)=\frac{n}{\sqrt{\pi}}\exp ^(-n^2x^2)\\
& \delta '_n(x)=-\frac{2n^3x}{\sqrt{\pi}}\exp ^(-n^2x^2),
\end{eqnarray}
where $n$ is a positive integer.
Using the relations $x\delta '(x)=-\delta (x)$ and $x^2\delta '(x)=-x\delta (x)=0$, 
we observe from Eq.~\eqref{2dd} 
\begin{multline}
(x-x_s)\frac{\partial^2 v_{(0)}}{\partial x^2}\\
=(x-x_s)\left(1-\Theta (x-x_s)\right)\frac{d^2f_1}{dx^2}+(x-x_s)\Theta (x-x_s)\frac{d^2f_2}{dx^2}\\
-\delta (x-x_s)\left(f_2(x)-f_1(x)\right). 
\end{multline}
Then it follows that 
\begin{eqnarray}
(x-x_s)^n\frac{\partial^2 v_{(0)}}{\partial x^2}|_{x=x_s}=-\delta_{n,1}\delta (0)\left(v_2-v_1\right).
\end{eqnarray}
This is how ``strange" the second derivative 
$\frac{\partial^2 v_{(0)}}{\partial x^2}|_{x=x_s}$ in fact behaves. 

\section{Initiation of oscillations in dispersive shock waves}\label{numerics} 
\begin{figure}[hbt]
\centering
\includegraphics[scale=0.22]{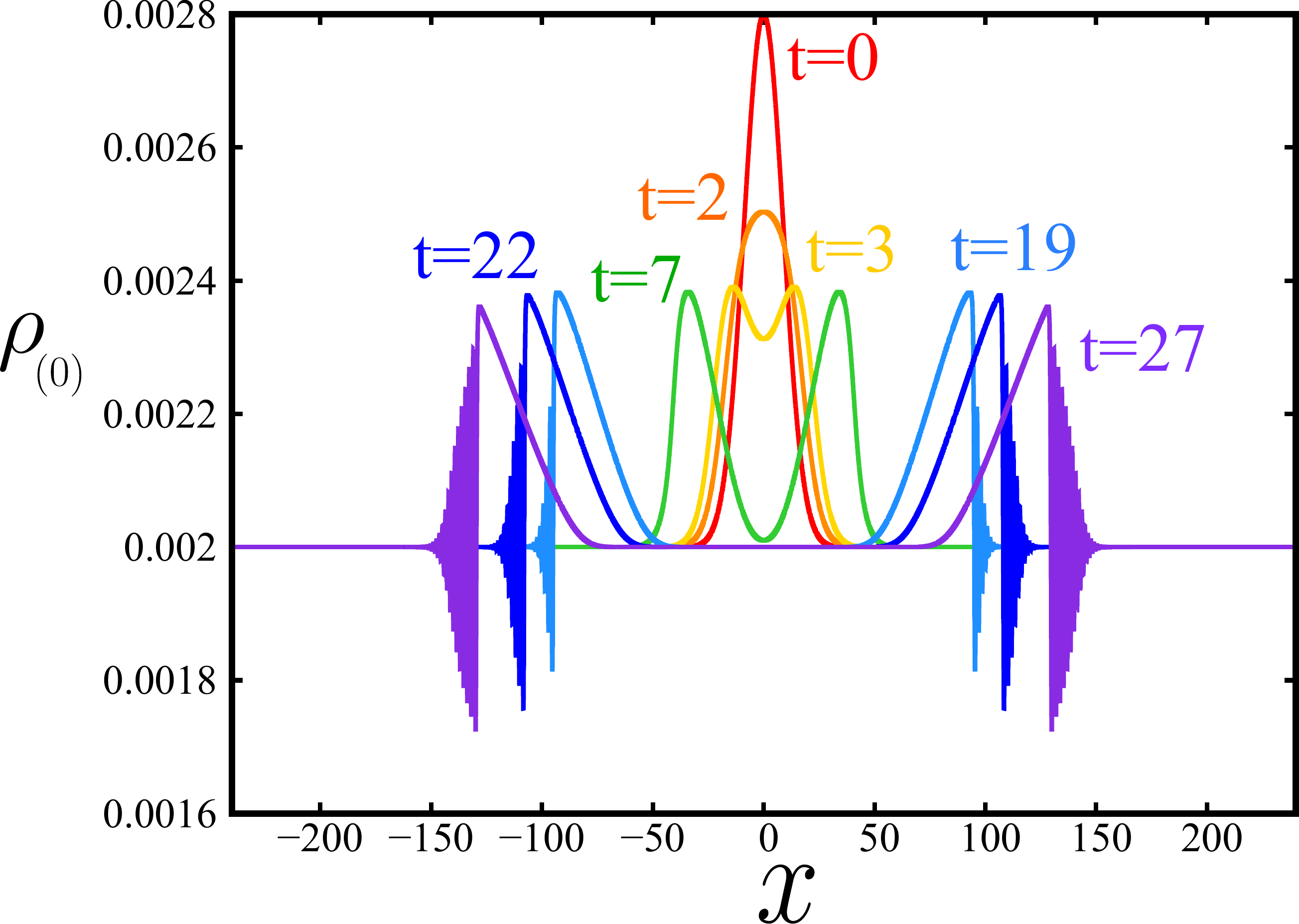}
\includegraphics[scale=0.22]{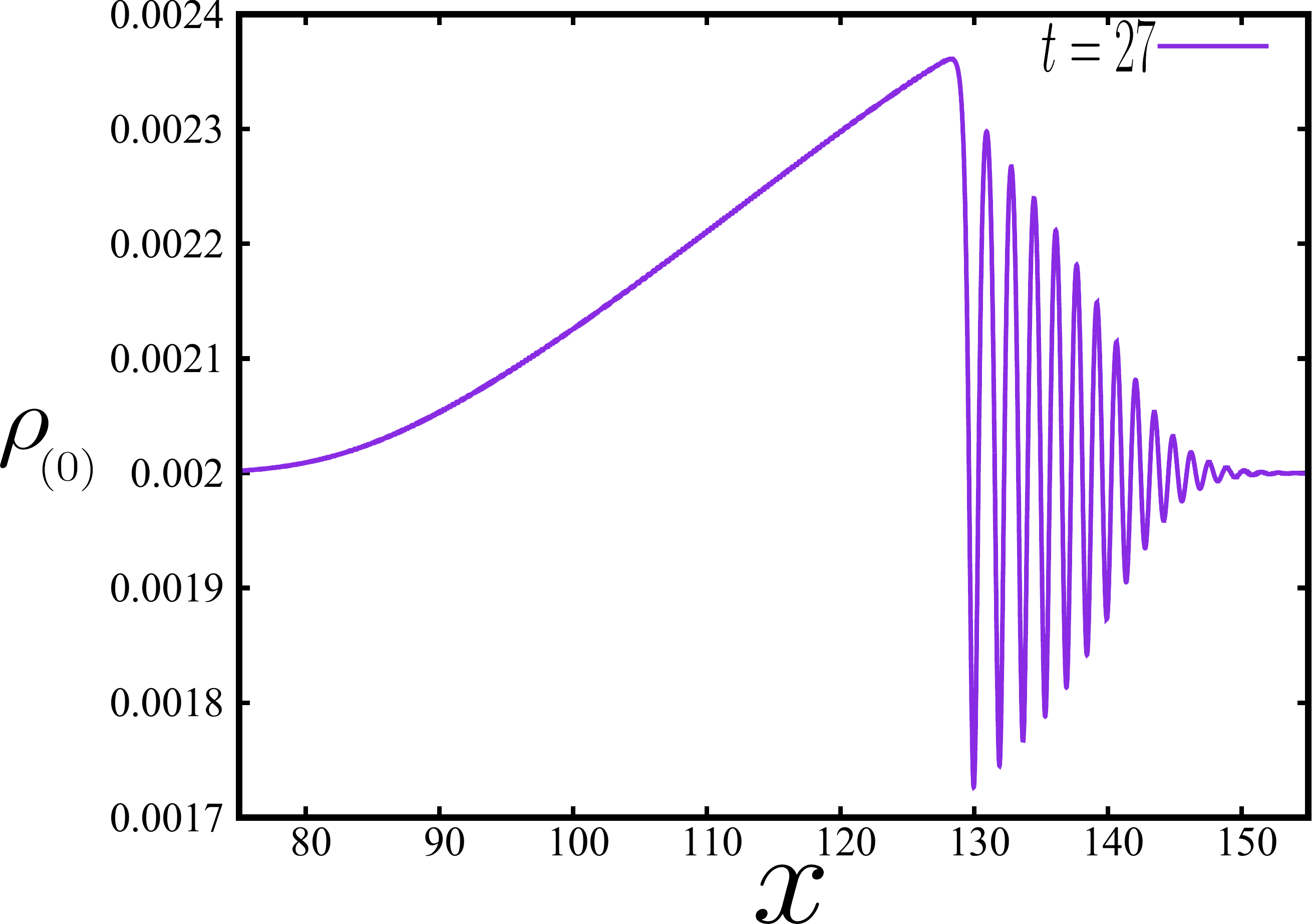}
  \caption{(Top) Evolution of density profile with time. At $t=0$, the laser at the center of the condensate is switched off. The initial Gaussian density profile splits in two parts, moving in opposite directions, and an 
  oscillation pattern is created, as described in Ref.~\cite{Damski2004}. (Bottom) Zoomed-in view of the density profile in the oscillation region at $t=27$.  Parameters as in Fig.~\ref{fig1}.}

  
\label{fig4}
\end{figure}
In this Appendix, we collect our numerical findings {on dispersive shock waves with initial 
conditions \eqref{roin}, as described in the main text.} {Some of these results have been presented already 
in Ref.~\cite{Damski2004}, but for the convenience of the reader we reproduce here these results together with a 
few additional observations, where our overall aim is to inspect closely the initiation of the oscillation of the dispersive shock waves, which is due to the quantum pressure term.} 

\begin{figure}[t]
\centering
\includegraphics[scale=0.21]{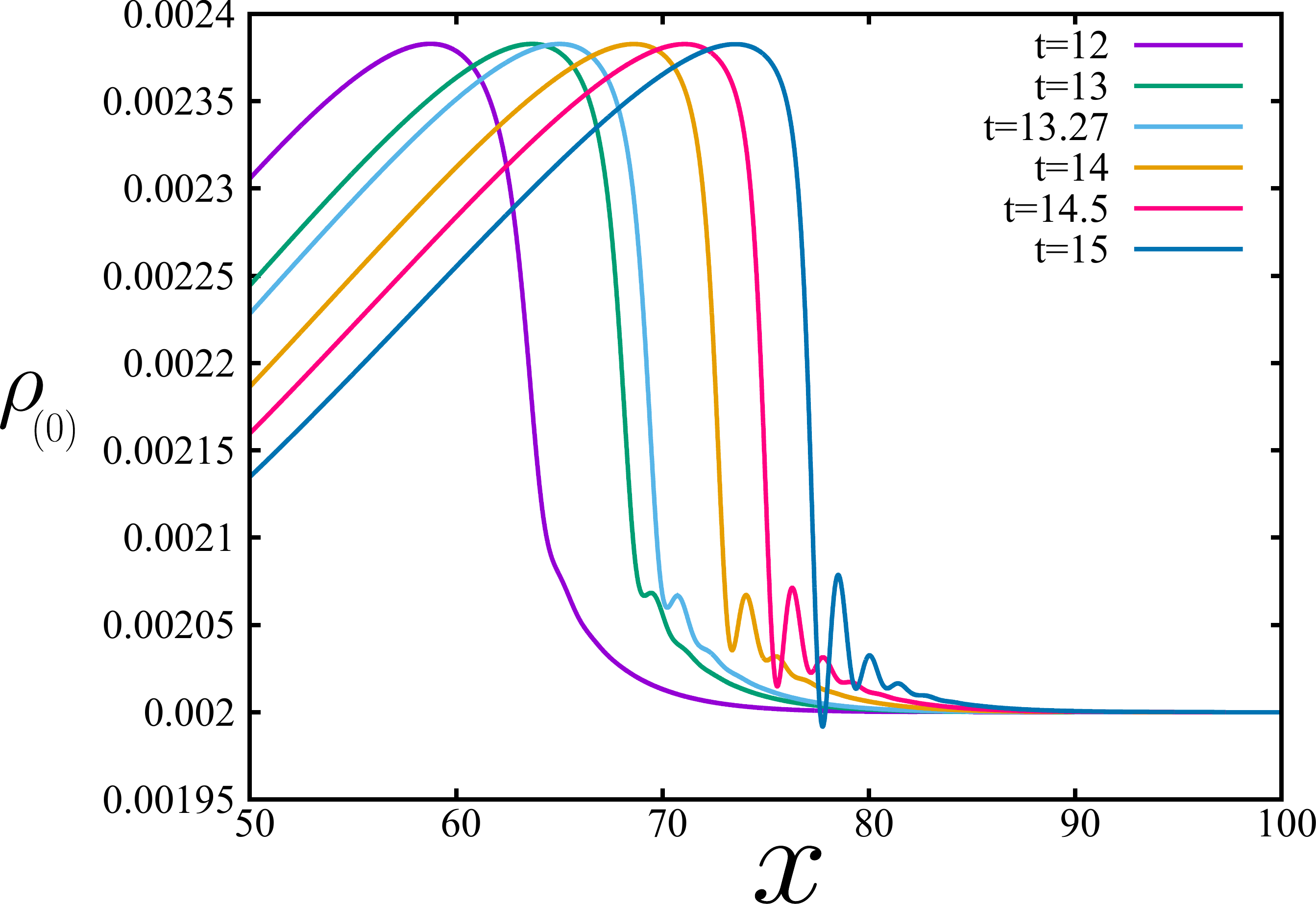}
\caption{Density near the shock time, $t_{\rm shock}\simeq 13.43$. An oscillation starts instead of a discontinuity 
popping up, due to the presence of quantum pressure. The initial wave parameters are as in Fig.~\ref{fig1}.}
\label{fig7}
\end{figure}

\begin{figure}[hbt]
\centering
\subfigure[]{\includegraphics[width=0.237\textwidth]{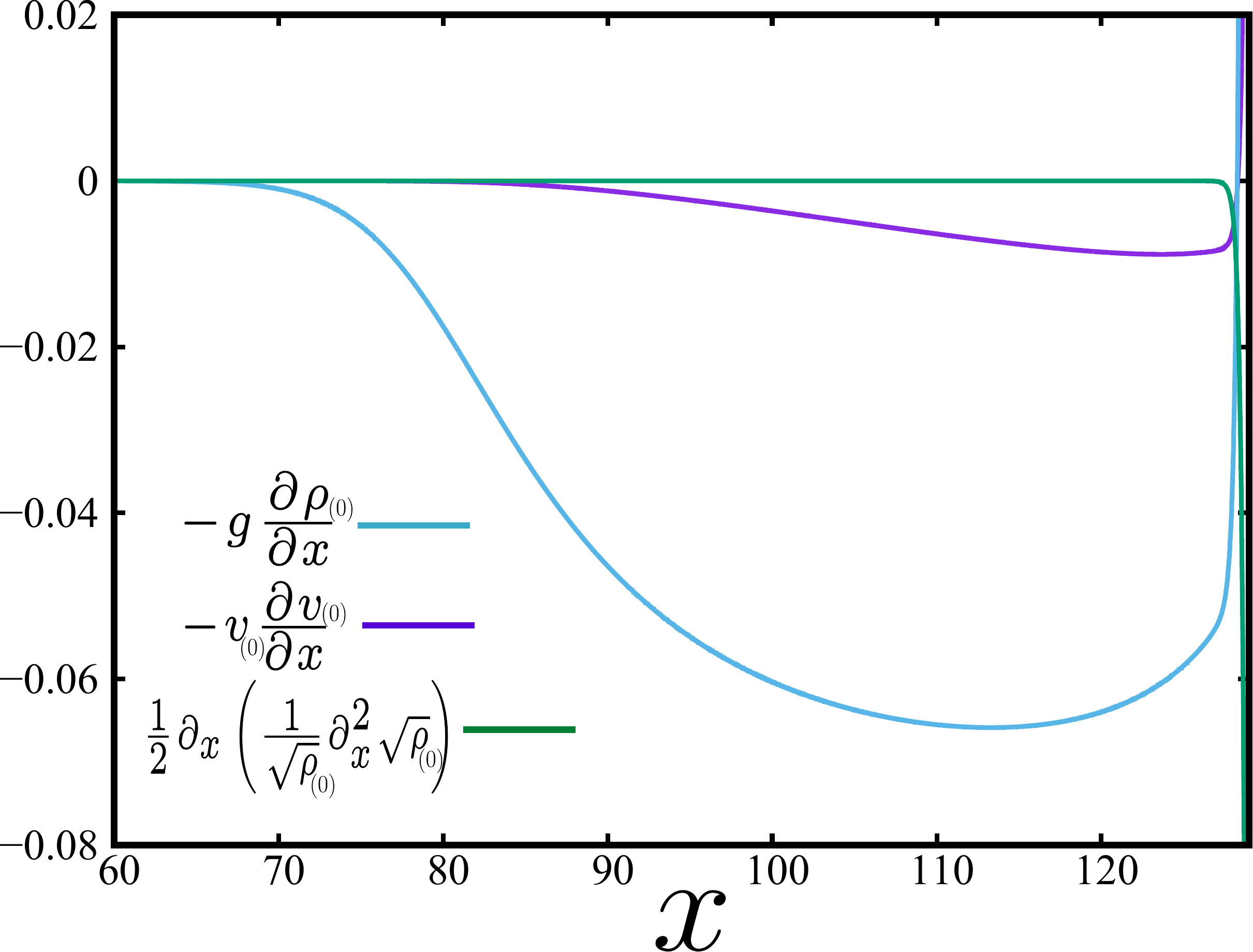}}
\subfigure[]{\includegraphics[width=0.24\textwidth]{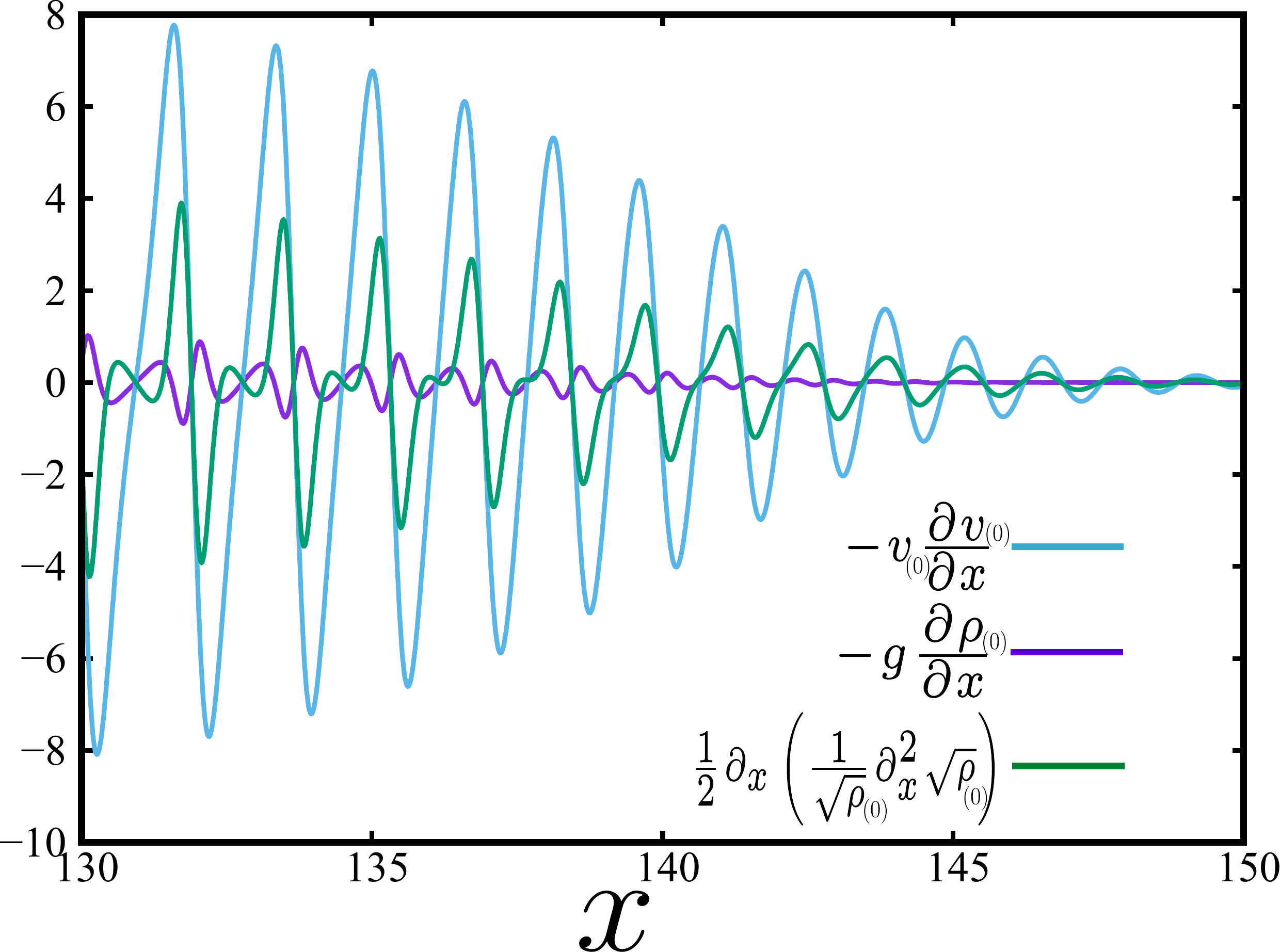}}

\subfigure[]{\includegraphics[width=0.285\textwidth]{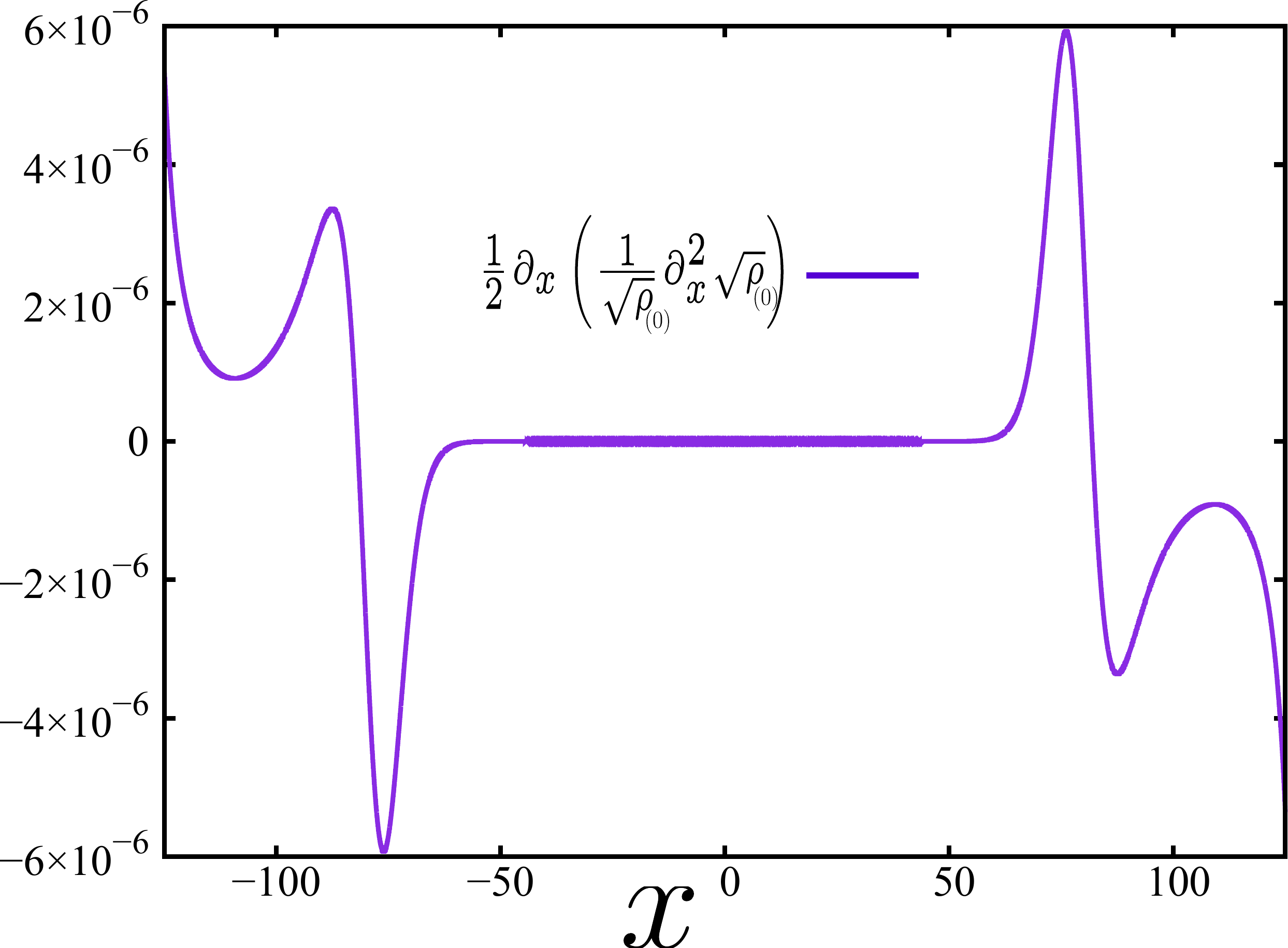}}
     \caption{(Fig. (a) and Fig. (b)) We compare the contribution of the terms involving derivatives in $x$ in the 1D momentum equation Eq. \eqref{euler} at $t=27$. We see that the quantum pressure term becomes significant only in the oscillatory region. 
     (Fig. (c)) The quantum pressure is very much smaller, by several orders of magnitudes, in the nonoscillatory region. The parameters are identical to those in Fig.~\ref{fig1}.}
\label{fig5}  
\end{figure}

Specifically, in Fig.~\ref{fig4}, 
we observe how the oscillation region is slowly spreading with progressing time. 
{In Fig.~\ref{fig7}, we display how the shock wave enters the oscillation phase, just prior to the shock time $t_{\rm shock}$.} Finally, in Fig.~\ref{fig5}, we display in some detail the onset of oscillations due to the quantum pressure becoming significant.

\vspace{0.5em}
\newpage
\bibliography{naked24}

\begin{thebibliography}{88}%
\makeatletter
\providecommand \@ifxundefined [1]{%
 \@ifx{#1\undefined}
}%
\providecommand \@ifnum [1]{%
 \ifnum #1\expandafter \@firstoftwo
 \else \expandafter \@secondoftwo
 \fi
}%
\providecommand \@ifx [1]{%
 \ifx #1\expandafter \@firstoftwo
 \else \expandafter \@secondoftwo
 \fi
}%
\providecommand \natexlab [1]{#1}%
\providecommand \enquote  [1]{``#1''}%
\providecommand \bibnamefont  [1]{#1}%
\providecommand \bibfnamefont [1]{#1}%
\providecommand \citenamefont [1]{#1}%
\providecommand \href@noop [0]{\@secondoftwo}%
\providecommand \href [0]{\begingroup \@sanitize@url \@href}%
\providecommand \@href[1]{\@@startlink{#1}\@@href}%
\providecommand \@@href[1]{\endgroup#1\@@endlink}%
\providecommand \@sanitize@url [0]{\catcode `\\12\catcode `\$12\catcode
  `\&12\catcode `\#12\catcode `\^12\catcode `\_12\catcode `\%12\relax}%
\providecommand \@@startlink[1]{}%
\providecommand \@@endlink[0]{}%
\providecommand \url  [0]{\begingroup\@sanitize@url \@url }%
\providecommand \@url [1]{\endgroup\@href {#1}{\urlprefix }}%
\providecommand \urlprefix  [0]{URL }%
\providecommand \Eprint [0]{\href }%
\providecommand \doibase [0]{https://doi.org/}%
\providecommand \selectlanguage [0]{\@gobble}%
\providecommand \bibinfo  [0]{\@secondoftwo}%
\providecommand \bibfield  [0]{\@secondoftwo}%
\providecommand \translation [1]{[#1]}%
\providecommand \BibitemOpen [0]{}%
\providecommand \bibitemStop [0]{}%
\providecommand \bibitemNoStop [0]{.\EOS\space}%
\providecommand \EOS [0]{\spacefactor3000\relax}%
\providecommand \BibitemShut  [1]{\csname bibitem#1\endcsname}%
\let\auto@bib@innerbib\@empty
\bibitem [{\citenamefont {Penrose}(1965)}]{Penrose65PRL}%
  \BibitemOpen
  \bibfield  {author} {\bibinfo {author} {\bibfnamefont {R.}~\bibnamefont
  {Penrose}},\ }\bibfield  {title} {\bibinfo {title} {{Gravitational Collapse
  and Space-Time Singularities}},\ }\href
  {https://doi.org/10.1103/PhysRevLett.14.57} {\bibfield  {journal} {\bibinfo
  {journal} {Phys. Rev. Lett.}\ }\textbf {\bibinfo {volume} {14}},\ \bibinfo
  {pages} {57} (\bibinfo {year} {1965})}\BibitemShut {NoStop}%
\bibitem [{\citenamefont {Penrose}(1969)}]{Penrose}%
  \BibitemOpen
  \bibfield  {author} {\bibinfo {author} {\bibfnamefont {R.}~\bibnamefont
  {Penrose}},\ }\bibfield  {title} {\bibinfo {title} {{{Gravitational Collapse:
  the Role of General Relativity}}},\ }\href
  {https://doi.org/10.1023/A:1016578408204} {\bibfield  {journal} {\bibinfo
  {journal} {Riv. Nuovo Cim.}\ }\textbf {\bibinfo {volume} {1}},\ \bibinfo
  {pages} {252} (\bibinfo {year} {1969})}\BibitemShut {NoStop}%
\bibitem [{\citenamefont {Hawking}(1976)}]{HawkingPredict}%
  \BibitemOpen
  \bibfield  {author} {\bibinfo {author} {\bibfnamefont {S.~W.}\ \bibnamefont
  {Hawking}},\ }\bibfield  {title} {\bibinfo {title} {Breakdown of
  predictability in gravitational collapse},\ }\href
  {https://doi.org/10.1103/PhysRevD.14.2460} {\bibfield  {journal} {\bibinfo
  {journal} {Phys. Rev. D}\ }\textbf {\bibinfo {volume} {14}},\ \bibinfo
  {pages} {2460} (\bibinfo {year} {1976})}\BibitemShut {NoStop}%
\bibitem [{\citenamefont {Hawking}\ and\ \citenamefont
  {Bondi}(1966{\natexlab{a}})}]{doi:10.1098/rspa.1966.0221}%
  \BibitemOpen
  \bibfield  {author} {\bibinfo {author} {\bibfnamefont {S.~W.}\ \bibnamefont
  {Hawking}}\ and\ \bibinfo {author} {\bibfnamefont {H.}~\bibnamefont
  {Bondi}},\ }\bibfield  {title} {\bibinfo {title} {The occurrence of
  singularities in cosmology},\ }\href {https://doi.org/10.1098/rspa.1966.0221}
  {\bibfield  {journal} {\bibinfo  {journal} {Proceedings of the Royal Society
  of London. Series A. Mathematical and Physical Sciences}\ }\textbf {\bibinfo
  {volume} {294}},\ \bibinfo {pages} {511} (\bibinfo {year}
  {1966}{\natexlab{a}})}\BibitemShut {NoStop}%
\bibitem [{\citenamefont {Hawking}\ and\ \citenamefont
  {Bondi}(1966{\natexlab{b}})}]{cs2}%
  \BibitemOpen
  \bibfield  {author} {\bibinfo {author} {\bibfnamefont {S.~W.}\ \bibnamefont
  {Hawking}}\ and\ \bibinfo {author} {\bibfnamefont {H.}~\bibnamefont
  {Bondi}},\ }\bibfield  {title} {\bibinfo {title} {The occurrence of
  singularities in cosmology. ii},\ }\href
  {https://doi.org/10.1098/rspa.1966.0255} {\bibfield  {journal} {\bibinfo
  {journal} {Proceedings of the Royal Society of London. Series A. Mathematical
  and Physical Sciences}\ }\textbf {\bibinfo {volume} {295}},\ \bibinfo {pages}
  {490} (\bibinfo {year} {1966}{\natexlab{b}})}\BibitemShut {NoStop}%
\bibitem [{\citenamefont {Hawking}(1967)}]{10.2307/2415769}%
  \BibitemOpen
  \bibfield  {author} {\bibinfo {author} {\bibfnamefont {S.~W.}\ \bibnamefont
  {Hawking}},\ }\bibfield  {title} {\bibinfo {title} {{The Occurrence of
  Singularities in Cosmology. III. Causality and Singularities}},\ }\href
  {http://www.jstor.org/stable/2415769} {\bibfield  {journal} {\bibinfo
  {journal} {Proceedings of the Royal Society of London. Series A, Mathematical
  and Physical Sciences}\ }\textbf {\bibinfo {volume} {300}},\ \bibinfo {pages}
  {187} (\bibinfo {year} {1967})}\BibitemShut {NoStop}%
\bibitem [{\citenamefont {Hawking}\ and\ \citenamefont
  {Penrose}(1970)}]{hawking1970singularities}%
  \BibitemOpen
  \bibfield  {author} {\bibinfo {author} {\bibfnamefont {S.~W.}\ \bibnamefont
  {Hawking}}\ and\ \bibinfo {author} {\bibfnamefont {R.}~\bibnamefont
  {Penrose}},\ }\bibfield  {title} {\bibinfo {title} {The singularities of
  gravitational collapse and cosmology},\ }\href
  {https://doi.org/10.1098/rspa.1970.0021} {\bibfield  {journal} {\bibinfo
  {journal} {Proceedings of the Royal Society of London. A. Mathematical and
  Physical Sciences}\ }\textbf {\bibinfo {volume} {314}},\ \bibinfo {pages}
  {529} (\bibinfo {year} {1970})}\BibitemShut {NoStop}%
\bibitem [{\citenamefont {Penrose}(1999)}]{Penrose99}%
  \BibitemOpen
  \bibfield  {author} {\bibinfo {author} {\bibfnamefont {R.}~\bibnamefont
  {Penrose}},\ }\bibfield  {title} {\bibinfo {title} {The question of cosmic
  censorship},\ }\href {https://doi.org/10.1007/BF02702355} {\bibfield
  {journal} {\bibinfo  {journal} {Journal of Astrophysics and Astronomy}\
  }\textbf {\bibinfo {volume} {20}},\ \bibinfo {pages} {233 } (\bibinfo {year}
  {1999})}\BibitemShut {NoStop}%
\bibitem [{\citenamefont {Wald}(1999)}]{Wald99}%
  \BibitemOpen
  \bibfield  {author} {\bibinfo {author} {\bibfnamefont {R.~M.}\ \bibnamefont
  {Wald}},\ }\bibinfo {title} {Gravitational collapse and cosmic censorship},\
  in\ \href {https://doi.org/10.1007/978-94-017-0934-7_5} {\emph {\bibinfo
  {booktitle} {Black Holes, Gravitational Radiation and the Universe: Essays in
  Honor of C.V. Vishveshwara}}},\ \bibinfo {editor} {edited by\ \bibinfo
  {editor} {\bibfnamefont {B.~R.}\ \bibnamefont {Iyer}}\ and\ \bibinfo {editor}
  {\bibfnamefont {B.}~\bibnamefont {Bhawal}}}\ (\bibinfo  {publisher} {Springer
  Netherlands},\ \bibinfo {address} {Dordrecht},\ \bibinfo {year} {1999})\ pp.\
  \bibinfo {pages} {69--86}\BibitemShut {NoStop}%
\bibitem [{\citenamefont {Christodoulou}(1984)}]{Christodoulou1984}%
  \BibitemOpen
  \bibfield  {author} {\bibinfo {author} {\bibfnamefont {D.}~\bibnamefont
  {Christodoulou}},\ }\bibfield  {title} {\bibinfo {title} {Violation of cosmic
  censorship in the gravitational collapse of a dust cloud},\ }\href
  {https://doi.org/10.1007/BF01223743} {\bibfield  {journal} {\bibinfo
  {journal} {Communications in Mathematical Physics}\ }\textbf {\bibinfo
  {volume} {93}},\ \bibinfo {pages} {171 } (\bibinfo {year}
  {1984})}\BibitemShut {NoStop}%
\bibitem [{\citenamefont {Roberts}(1989)}]{Roberts}%
  \BibitemOpen
  \bibfield  {author} {\bibinfo {author} {\bibfnamefont {M.~D.}\ \bibnamefont
  {Roberts}},\ }\bibfield  {title} {\bibinfo {title} {Scalar field
  counterexamples to the cosmic censorship hypothesis},\ }\href
  {https://doi.org/10.1007/BF00769864} {\bibfield  {journal} {\bibinfo
  {journal} {General Relativity and Gravitation}\ }\textbf {\bibinfo {volume}
  {21}},\ \bibinfo {pages} {907} (\bibinfo {year} {1989})}\BibitemShut
  {NoStop}%
\bibitem [{\citenamefont {Hubeny}(1999)}]{PhysRevD.59.064013}%
  \BibitemOpen
  \bibfield  {author} {\bibinfo {author} {\bibfnamefont {V.~E.}\ \bibnamefont
  {Hubeny}},\ }\bibfield  {title} {\bibinfo {title} {Overcharging a black hole
  and cosmic censorship},\ }\href {https://doi.org/10.1103/PhysRevD.59.064013}
  {\bibfield  {journal} {\bibinfo  {journal} {Phys. Rev. D}\ }\textbf {\bibinfo
  {volume} {59}},\ \bibinfo {pages} {064013} (\bibinfo {year}
  {1999})}\BibitemShut {NoStop}%
\bibitem [{\citenamefont {Matsas}\ and\ \citenamefont
  {da~Silva}(2007)}]{PhysRevLett.99.181301}%
  \BibitemOpen
  \bibfield  {author} {\bibinfo {author} {\bibfnamefont {G.~E.~A.}\
  \bibnamefont {Matsas}}\ and\ \bibinfo {author} {\bibfnamefont {A.~R.~R.}\
  \bibnamefont {da~Silva}},\ }\bibfield  {title} {\bibinfo {title}
  {{Overspinning a Nearly Extreme Charged Black Hole via a Quantum Tunneling
  Process}},\ }\href {https://doi.org/10.1103/PhysRevLett.99.181301} {\bibfield
   {journal} {\bibinfo  {journal} {Phys. Rev. Lett.}\ }\textbf {\bibinfo
  {volume} {99}},\ \bibinfo {pages} {181301} (\bibinfo {year}
  {2007})}\BibitemShut {NoStop}%
\bibitem [{\citenamefont {Matsas}\ \emph {et~al.}(2009)\citenamefont {Matsas},
  \citenamefont {Richartz}, \citenamefont {Saa}, \citenamefont {da~Silva},\
  and\ \citenamefont {Vanzella}}]{PhysRevD.79.101502}%
  \BibitemOpen
  \bibfield  {author} {\bibinfo {author} {\bibfnamefont {G.~E.~A.}\
  \bibnamefont {Matsas}}, \bibinfo {author} {\bibfnamefont {M.}~\bibnamefont
  {Richartz}}, \bibinfo {author} {\bibfnamefont {A.}~\bibnamefont {Saa}},
  \bibinfo {author} {\bibfnamefont {A.~R.~R.}\ \bibnamefont {da~Silva}},\ and\
  \bibinfo {author} {\bibfnamefont {D.~A.~T.}\ \bibnamefont {Vanzella}},\
  }\bibfield  {title} {\bibinfo {title} {Can quantum mechanics fool the cosmic
  censor?},\ }\href {https://doi.org/10.1103/PhysRevD.79.101502} {\bibfield
  {journal} {\bibinfo  {journal} {Phys. Rev. D}\ }\textbf {\bibinfo {volume}
  {79}},\ \bibinfo {pages} {101502(R)} (\bibinfo {year} {2009})}\BibitemShut
  {NoStop}%
\bibitem [{\citenamefont {Hod}(2008{\natexlab{a}})}]{Hod}%
  \BibitemOpen
  \bibfield  {author} {\bibinfo {author} {\bibfnamefont {S.}~\bibnamefont
  {Hod}},\ }\bibfield  {title} {\bibinfo {title} {{Weak Cosmic Censorship: As
  Strong as Ever}},\ }\href {https://doi.org/10.1103/PhysRevLett.100.121101}
  {\bibfield  {journal} {\bibinfo  {journal} {Phys. Rev. Lett.}\ }\textbf
  {\bibinfo {volume} {100}},\ \bibinfo {pages} {121101} (\bibinfo {year}
  {2008}{\natexlab{a}})}\BibitemShut {NoStop}%
\bibitem [{\citenamefont {Casals}\ \emph {et~al.}(2016)\citenamefont {Casals},
  \citenamefont {Fabbri}, \citenamefont {Martínez},\ and\ \citenamefont
  {Zanelli}}]{CASALS2016244}%
  \BibitemOpen
  \bibfield  {author} {\bibinfo {author} {\bibfnamefont {M.}~\bibnamefont
  {Casals}}, \bibinfo {author} {\bibfnamefont {A.}~\bibnamefont {Fabbri}},
  \bibinfo {author} {\bibfnamefont {C.}~\bibnamefont {Martínez}},\ and\
  \bibinfo {author} {\bibfnamefont {J.}~\bibnamefont {Zanelli}},\ }\bibfield
  {title} {\bibinfo {title} {Quantum dress for a naked singularity},\ }\href
  {https://doi.org/https://doi.org/10.1016/j.physletb.2016.06.044} {\bibfield
  {journal} {\bibinfo  {journal} {Physics Letters B}\ }\textbf {\bibinfo
  {volume} {760}},\ \bibinfo {pages} {244} (\bibinfo {year}
  {2016})}\BibitemShut {NoStop}%
\bibitem [{\citenamefont {{Wald}}(1974)}]{1974AnPhy..82..548W}%
  \BibitemOpen
  \bibfield  {author} {\bibinfo {author} {\bibfnamefont {R.}~\bibnamefont
  {{Wald}}},\ }\bibfield  {title} {\bibinfo {title} {{Gedanken experiments to
  destroy a black hole.}},\ }\href
  {https://doi.org/10.1016/0003-4916(74)90125-0} {\bibfield  {journal}
  {\bibinfo  {journal} {Annals of Physics}\ }\textbf {\bibinfo {volume} {82}},\
  \bibinfo {pages} {548} (\bibinfo {year} {1974})}\BibitemShut {NoStop}%
\bibitem [{\citenamefont {Sorce}\ and\ \citenamefont
  {Wald}(2017)}]{PhysRevD.96.104014}%
  \BibitemOpen
  \bibfield  {author} {\bibinfo {author} {\bibfnamefont {J.}~\bibnamefont
  {Sorce}}\ and\ \bibinfo {author} {\bibfnamefont {R.~M.}\ \bibnamefont
  {Wald}},\ }\bibfield  {title} {\bibinfo {title} {{Gedanken experiments to
  destroy a black hole. II. Kerr-Newman black holes cannot be overcharged or
  overspun}},\ }\href {https://doi.org/10.1103/PhysRevD.96.104014} {\bibfield
  {journal} {\bibinfo  {journal} {Phys. Rev. D}\ }\textbf {\bibinfo {volume}
  {96}},\ \bibinfo {pages} {104014} (\bibinfo {year} {2017})}\BibitemShut
  {NoStop}%
\bibitem [{\citenamefont {Hod}(2008{\natexlab{b}})}]{Hod2008}%
  \BibitemOpen
  \bibfield  {author} {\bibinfo {author} {\bibfnamefont {S.}~\bibnamefont
  {Hod}},\ }\bibfield  {title} {\bibinfo {title} {Return of the quantum cosmic
  censor},\ }\href
  {https://doi.org/https://doi.org/10.1016/j.physletb.2008.08.059} {\bibfield
  {journal} {\bibinfo  {journal} {Physics Letters B}\ }\textbf {\bibinfo
  {volume} {668}},\ \bibinfo {pages} {346} (\bibinfo {year}
  {2008}{\natexlab{b}})}\BibitemShut {NoStop}%
\bibitem [{\citenamefont {Unruh}(1981)}]{unruh}%
  \BibitemOpen
  \bibfield  {author} {\bibinfo {author} {\bibfnamefont {W.~G.}\ \bibnamefont
  {Unruh}},\ }\bibfield  {title} {\bibinfo {title} {{Experimental Black-Hole
  Evaporation?}},\ }\href {https://doi.org/10.1103/PhysRevLett.46.1351}
  {\bibfield  {journal} {\bibinfo  {journal} {Phys. Rev. Lett.}\ }\textbf
  {\bibinfo {volume} {46}},\ \bibinfo {pages} {1351} (\bibinfo {year}
  {1981})}\BibitemShut {NoStop}%
\bibitem [{\citenamefont {Barcel{\'o}}\ \emph {et~al.}(2011)\citenamefont
  {Barcel{\'o}}, \citenamefont {Liberati},\ and\ \citenamefont {Visser}}]{BLV}%
  \BibitemOpen
  \bibfield  {author} {\bibinfo {author} {\bibfnamefont {C.}~\bibnamefont
  {Barcel{\'o}}}, \bibinfo {author} {\bibfnamefont {S.}~\bibnamefont
  {Liberati}},\ and\ \bibinfo {author} {\bibfnamefont {M.}~\bibnamefont
  {Visser}},\ }\bibfield  {title} {\bibinfo {title} {{Analogue Gravity}},\
  }\href {https://doi.org/10.12942/lrr-2011-3} {\bibfield  {journal} {\bibinfo
  {journal} {Living Reviews in Relativity}\ }\textbf {\bibinfo {volume} {14}},\
  \bibinfo {pages} {3} (\bibinfo {year} {2011})}\BibitemShut {NoStop}%
\bibitem [{\citenamefont {Barcel{\'{o}}}\ \emph
  {et~al.}(2001{\natexlab{a}})\citenamefont {Barcel{\'{o}}}, \citenamefont
  {Liberati},\ and\ \citenamefont {Visser}}]{BLV_normal}%
  \BibitemOpen
  \bibfield  {author} {\bibinfo {author} {\bibfnamefont {C.}~\bibnamefont
  {Barcel{\'{o}}}}, \bibinfo {author} {\bibfnamefont {S.}~\bibnamefont
  {Liberati}},\ and\ \bibinfo {author} {\bibfnamefont {M.}~\bibnamefont
  {Visser}},\ }\bibfield  {title} {\bibinfo {title} {Analogue gravity from
  field theory normal modes?},\ }\href
  {https://doi.org/10.1088/0264-9381/18/17/313} {\bibfield  {journal} {\bibinfo
   {journal} {Classical and Quantum Gravity}\ }\textbf {\bibinfo {volume}
  {18}},\ \bibinfo {pages} {3595} (\bibinfo {year}
  {2001}{\natexlab{a}})}\BibitemShut {NoStop}%
\bibitem [{\citenamefont {Sch\"utzhold}\ and\ \citenamefont
  {Unruh}(2002)}]{RalfBill}%
  \BibitemOpen
  \bibfield  {author} {\bibinfo {author} {\bibfnamefont {R.}~\bibnamefont
  {Sch\"utzhold}}\ and\ \bibinfo {author} {\bibfnamefont {W.~G.}\ \bibnamefont
  {Unruh}},\ }\bibfield  {title} {\bibinfo {title} {Gravity wave analogues of
  black holes},\ }\href {https://doi.org/10.1103/PhysRevD.66.044019} {\bibfield
   {journal} {\bibinfo  {journal} {Phys. Rev. D}\ }\textbf {\bibinfo {volume}
  {66}},\ \bibinfo {pages} {044019} (\bibinfo {year} {2002})}\BibitemShut
  {NoStop}%
\bibitem [{\citenamefont {Weinfurtner}\ \emph {et~al.}(2011)\citenamefont
  {Weinfurtner}, \citenamefont {Tedford}, \citenamefont {Penrice},
  \citenamefont {Unruh},\ and\ \citenamefont {Lawrence}}]{Weinfurtner}%
  \BibitemOpen
  \bibfield  {author} {\bibinfo {author} {\bibfnamefont {S.}~\bibnamefont
  {Weinfurtner}}, \bibinfo {author} {\bibfnamefont {E.~W.}\ \bibnamefont
  {Tedford}}, \bibinfo {author} {\bibfnamefont {M.~C.~J.}\ \bibnamefont
  {Penrice}}, \bibinfo {author} {\bibfnamefont {W.~G.}\ \bibnamefont {Unruh}},\
  and\ \bibinfo {author} {\bibfnamefont {G.~A.}\ \bibnamefont {Lawrence}},\
  }\bibfield  {title} {\bibinfo {title} {{Measurement of Stimulated Hawking
  Emission in an Analogue System}},\ }\href
  {https://doi.org/10.1103/PhysRevLett.106.021302} {\bibfield  {journal}
  {\bibinfo  {journal} {Phys. Rev. Lett.}\ }\textbf {\bibinfo {volume} {106}},\
  \bibinfo {pages} {021302} (\bibinfo {year} {2011})}\BibitemShut {NoStop}%
\bibitem [{\citenamefont {Euv\'e}\ \emph {et~al.}(2016)\citenamefont {Euv\'e},
  \citenamefont {Michel}, \citenamefont {Parentani}, \citenamefont {Philbin},\
  and\ \citenamefont {Rousseaux}}]{Euve}%
  \BibitemOpen
  \bibfield  {author} {\bibinfo {author} {\bibfnamefont {L.-P.}\ \bibnamefont
  {Euv\'e}}, \bibinfo {author} {\bibfnamefont {F.}~\bibnamefont {Michel}},
  \bibinfo {author} {\bibfnamefont {R.}~\bibnamefont {Parentani}}, \bibinfo
  {author} {\bibfnamefont {T.~G.}\ \bibnamefont {Philbin}},\ and\ \bibinfo
  {author} {\bibfnamefont {G.}~\bibnamefont {Rousseaux}},\ }\bibfield  {title}
  {\bibinfo {title} {{Observation of Noise Correlated by the Hawking Effect in
  a Water Tank}},\ }\href {https://doi.org/10.1103/PhysRevLett.117.121301}
  {\bibfield  {journal} {\bibinfo  {journal} {Phys. Rev. Lett.}\ }\textbf
  {\bibinfo {volume} {117}},\ \bibinfo {pages} {121301} (\bibinfo {year}
  {2016})}\BibitemShut {NoStop}%
\bibitem [{\citenamefont {Euv\'e}\ \emph
  {et~al.}(2020{\natexlab{a}})\citenamefont {Euv\'e}, \citenamefont
  {Robertson}, \citenamefont {James}, \citenamefont {Fabbri},\ and\
  \citenamefont {Rousseaux}}]{EuveII}%
  \BibitemOpen
  \bibfield  {author} {\bibinfo {author} {\bibfnamefont {L.-P.}\ \bibnamefont
  {Euv\'e}}, \bibinfo {author} {\bibfnamefont {S.}~\bibnamefont {Robertson}},
  \bibinfo {author} {\bibfnamefont {N.}~\bibnamefont {James}}, \bibinfo
  {author} {\bibfnamefont {A.}~\bibnamefont {Fabbri}},\ and\ \bibinfo {author}
  {\bibfnamefont {G.}~\bibnamefont {Rousseaux}},\ }\bibfield  {title} {\bibinfo
  {title} {{Scattering of Co-Current Surface Waves on an Analogue Black
  Hole}},\ }\href {https://doi.org/10.1103/PhysRevLett.124.141101} {\bibfield
  {journal} {\bibinfo  {journal} {Phys. Rev. Lett.}\ }\textbf {\bibinfo
  {volume} {124}},\ \bibinfo {pages} {141101} (\bibinfo {year}
  {2020}{\natexlab{a}})}\BibitemShut {NoStop}%
\bibitem [{\citenamefont {Marino}(2008)}]{Marino}%
  \BibitemOpen
  \bibfield  {author} {\bibinfo {author} {\bibfnamefont {F.}~\bibnamefont
  {Marino}},\ }\bibfield  {title} {\bibinfo {title} {Acoustic black holes in a
  two-dimensional ``photon fluid''},\ }\href
  {https://doi.org/10.1103/PhysRevA.78.063804} {\bibfield  {journal} {\bibinfo
  {journal} {Phys. Rev. A}\ }\textbf {\bibinfo {volume} {78}},\ \bibinfo
  {pages} {063804} (\bibinfo {year} {2008})}\BibitemShut {NoStop}%
\bibitem [{\citenamefont {Nguyen}\ \emph {et~al.}(2015)\citenamefont {Nguyen},
  \citenamefont {Gerace}, \citenamefont {Carusotto}, \citenamefont {Sanvitto},
  \citenamefont {Galopin}, \citenamefont {Lema\^{\i}tre}, \citenamefont
  {Sagnes}, \citenamefont {Bloch},\ and\ \citenamefont {Amo}}]{Nguyen}%
  \BibitemOpen
  \bibfield  {author} {\bibinfo {author} {\bibfnamefont {H.~S.}\ \bibnamefont
  {Nguyen}}, \bibinfo {author} {\bibfnamefont {D.}~\bibnamefont {Gerace}},
  \bibinfo {author} {\bibfnamefont {I.}~\bibnamefont {Carusotto}}, \bibinfo
  {author} {\bibfnamefont {D.}~\bibnamefont {Sanvitto}}, \bibinfo {author}
  {\bibfnamefont {E.}~\bibnamefont {Galopin}}, \bibinfo {author} {\bibfnamefont
  {A.}~\bibnamefont {Lema\^{\i}tre}}, \bibinfo {author} {\bibfnamefont
  {I.}~\bibnamefont {Sagnes}}, \bibinfo {author} {\bibfnamefont
  {J.}~\bibnamefont {Bloch}},\ and\ \bibinfo {author} {\bibfnamefont
  {A.}~\bibnamefont {Amo}},\ }\bibfield  {title} {\bibinfo {title} {{Acoustic
  Black Hole in a Stationary Hydrodynamic Flow of Microcavity Polaritons}},\
  }\href {https://doi.org/10.1103/PhysRevLett.114.036402} {\bibfield  {journal}
  {\bibinfo  {journal} {Phys. Rev. Lett.}\ }\textbf {\bibinfo {volume} {114}},\
  \bibinfo {pages} {036402} (\bibinfo {year} {2015})}\BibitemShut {NoStop}%
\bibitem [{\citenamefont {Jacquet}\ \emph {et~al.}(2022)\citenamefont
  {Jacquet}, \citenamefont {Joly}, \citenamefont {Claude}, \citenamefont
  {Giacomelli}, \citenamefont {Glorieux}, \citenamefont {Bramati},
  \citenamefont {Carusotto},\ and\ \citenamefont {Giacobino}}]{polariton}%
  \BibitemOpen
  \bibfield  {author} {\bibinfo {author} {\bibfnamefont {M.}~\bibnamefont
  {Jacquet}}, \bibinfo {author} {\bibfnamefont {M.}~\bibnamefont {Joly}},
  \bibinfo {author} {\bibfnamefont {F.}~\bibnamefont {Claude}}, \bibinfo
  {author} {\bibfnamefont {L.}~\bibnamefont {Giacomelli}}, \bibinfo {author}
  {\bibfnamefont {Q.}~\bibnamefont {Glorieux}}, \bibinfo {author}
  {\bibfnamefont {A.}~\bibnamefont {Bramati}}, \bibinfo {author} {\bibfnamefont
  {I.}~\bibnamefont {Carusotto}},\ and\ \bibinfo {author} {\bibfnamefont
  {E.}~\bibnamefont {Giacobino}},\ }\bibfield  {title} {\bibinfo {title}
  {{Analogue quantum simulation of the Hawking effect in a polariton
  superfluid}},\ }\href {https://doi.org/10.1140/epjd/s10053-022-00477-5}
  {\bibfield  {journal} {\bibinfo  {journal} {The European Physical Journal D}\
  }\textbf {\bibinfo {volume} {76}},\ \bibinfo {pages} {152} (\bibinfo {year}
  {2022})}\BibitemShut {NoStop}%
\bibitem [{\citenamefont {Torres}\ \emph {et~al.}(2020)\citenamefont {Torres},
  \citenamefont {Patrick}, \citenamefont {Richartz},\ and\ \citenamefont
  {Weinfurtner}}]{Richartz}%
  \BibitemOpen
  \bibfield  {author} {\bibinfo {author} {\bibfnamefont {T.}~\bibnamefont
  {Torres}}, \bibinfo {author} {\bibfnamefont {S.}~\bibnamefont {Patrick}},
  \bibinfo {author} {\bibfnamefont {M.}~\bibnamefont {Richartz}},\ and\
  \bibinfo {author} {\bibfnamefont {S.}~\bibnamefont {Weinfurtner}},\
  }\bibfield  {title} {\bibinfo {title} {Quasinormal mode oscillations in an
  analogue black hole experiment},\ }\href
  {https://doi.org/10.1103/PhysRevLett.125.011301} {\bibfield  {journal}
  {\bibinfo  {journal} {Phys. Rev. Lett.}\ }\textbf {\bibinfo {volume} {125}},\
  \bibinfo {pages} {011301} (\bibinfo {year} {2020})}\BibitemShut {NoStop}%
\bibitem [{\citenamefont {Datta}(2018)}]{PhysRevD.98.064049}%
  \BibitemOpen
  \bibfield  {author} {\bibinfo {author} {\bibfnamefont {S.}~\bibnamefont
  {Datta}},\ }\bibfield  {title} {\bibinfo {title} {{Acoustic analog of
  gravitational wave}},\ }\href {https://doi.org/10.1103/PhysRevD.98.064049}
  {\bibfield  {journal} {\bibinfo  {journal} {Phys. Rev. D}\ }\textbf {\bibinfo
  {volume} {98}},\ \bibinfo {pages} {064049} (\bibinfo {year}
  {2018})}\BibitemShut {NoStop}%
\bibitem [{\citenamefont {Liberati}\ \emph {et~al.}(2019)\citenamefont
  {Liberati}, \citenamefont {Tricella},\ and\ \citenamefont
  {Trombettoni}}]{Andrea}%
  \BibitemOpen
  \bibfield  {author} {\bibinfo {author} {\bibfnamefont {S.}~\bibnamefont
  {Liberati}}, \bibinfo {author} {\bibfnamefont {G.}~\bibnamefont {Tricella}},\
  and\ \bibinfo {author} {\bibfnamefont {A.}~\bibnamefont {Trombettoni}},\
  }\bibfield  {title} {\bibinfo {title} {{The Information Loss Problem: An
  Analogue Gravity Perspective}},\ }\href {https://doi.org/10.3390/e21100940}
  {\bibfield  {journal} {\bibinfo  {journal} {Entropy}\ }\textbf {\bibinfo
  {volume} {21}},\ \bibinfo {pages} {940} (\bibinfo {year} {2019})}\BibitemShut
  {NoStop}%
\bibitem [{\citenamefont {Corley}\ and\ \citenamefont {Jacobson}(1999)}]{Ted}%
  \BibitemOpen
  \bibfield  {author} {\bibinfo {author} {\bibfnamefont {S.}~\bibnamefont
  {Corley}}\ and\ \bibinfo {author} {\bibfnamefont {T.}~\bibnamefont
  {Jacobson}},\ }\bibfield  {title} {\bibinfo {title} {Black hole lasers},\
  }\href {https://doi.org/10.1103/PhysRevD.59.124011} {\bibfield  {journal}
  {\bibinfo  {journal} {Phys. Rev. D}\ }\textbf {\bibinfo {volume} {59}},\
  \bibinfo {pages} {124011} (\bibinfo {year} {1999})}\BibitemShut {NoStop}%
\bibitem [{\citenamefont {Kosior}\ \emph {et~al.}(2018)\citenamefont {Kosior},
  \citenamefont {Lewenstein},\ and\ \citenamefont {Celi}}]{Celi}%
  \BibitemOpen
  \bibfield  {author} {\bibinfo {author} {\bibfnamefont {A.}~\bibnamefont
  {Kosior}}, \bibinfo {author} {\bibfnamefont {M.}~\bibnamefont {Lewenstein}},\
  and\ \bibinfo {author} {\bibfnamefont {A.}~\bibnamefont {Celi}},\ }\bibfield
  {title} {\bibinfo {title} {{Unruh effect for interacting particles with
  ultracold atoms}},\ }\href {https://doi.org/10.21468/SciPostPhys.5.6.061}
  {\bibfield  {journal} {\bibinfo  {journal} {SciPost Phys.}\ }\textbf
  {\bibinfo {volume} {5}},\ \bibinfo {pages} {61} (\bibinfo {year}
  {2018})}\BibitemShut {NoStop}%
\bibitem [{\citenamefont {Basak}\ and\ \citenamefont {Majumdar}(2003)}]{Basak}%
  \BibitemOpen
  \bibfield  {author} {\bibinfo {author} {\bibfnamefont {S.}~\bibnamefont
  {Basak}}\ and\ \bibinfo {author} {\bibfnamefont {P.}~\bibnamefont
  {Majumdar}},\ }\bibfield  {title} {\bibinfo {title} {{`Superresonance' from a
  rotating acoustic black hole}},\ }\href
  {https://doi.org/10.1088/0264-9381/20/18/304} {\bibfield  {journal} {\bibinfo
   {journal} {Classical and Quantum Gravity}\ }\textbf {\bibinfo {volume}
  {20}},\ \bibinfo {pages} {3907} (\bibinfo {year} {2003})}\BibitemShut
  {NoStop}%
\bibitem [{\citenamefont {Torres}\ \emph {et~al.}(2017)\citenamefont {Torres},
  \citenamefont {Patrick}, \citenamefont {Coutant}, \citenamefont {Richartz},
  \citenamefont {Tedford},\ and\ \citenamefont {Weinfurtner}}]{Torres}%
  \BibitemOpen
  \bibfield  {author} {\bibinfo {author} {\bibfnamefont {T.}~\bibnamefont
  {Torres}}, \bibinfo {author} {\bibfnamefont {S.}~\bibnamefont {Patrick}},
  \bibinfo {author} {\bibfnamefont {A.}~\bibnamefont {Coutant}}, \bibinfo
  {author} {\bibfnamefont {M.}~\bibnamefont {Richartz}}, \bibinfo {author}
  {\bibfnamefont {E.~W.}\ \bibnamefont {Tedford}},\ and\ \bibinfo {author}
  {\bibfnamefont {S.}~\bibnamefont {Weinfurtner}},\ }\bibfield  {title}
  {\bibinfo {title} {Rotational superradiant scattering in a vortex flow},\
  }\href {https://doi.org/10.1038/nphys4151} {\bibfield  {journal} {\bibinfo
  {journal} {Nature Physics}\ }\textbf {\bibinfo {volume} {13}},\ \bibinfo
  {pages} {833} (\bibinfo {year} {2017})}\BibitemShut {NoStop}%
\bibitem [{\citenamefont {Prain}\ \emph {et~al.}(2019)\citenamefont {Prain},
  \citenamefont {Maitland}, \citenamefont {Faccio},\ and\ \citenamefont
  {Marino}}]{Prain}%
  \BibitemOpen
  \bibfield  {author} {\bibinfo {author} {\bibfnamefont {A.}~\bibnamefont
  {Prain}}, \bibinfo {author} {\bibfnamefont {C.}~\bibnamefont {Maitland}},
  \bibinfo {author} {\bibfnamefont {D.}~\bibnamefont {Faccio}},\ and\ \bibinfo
  {author} {\bibfnamefont {F.}~\bibnamefont {Marino}},\ }\bibfield  {title}
  {\bibinfo {title} {Superradiant scattering in fluids of light},\ }\href
  {https://doi.org/10.1103/PhysRevD.100.024037} {\bibfield  {journal} {\bibinfo
   {journal} {Phys. Rev. D}\ }\textbf {\bibinfo {volume} {100}},\ \bibinfo
  {pages} {024037} (\bibinfo {year} {2019})}\BibitemShut {NoStop}%
\bibitem [{\citenamefont {Braidotti}\ \emph {et~al.}(2022)\citenamefont
  {Braidotti}, \citenamefont {Prizia}, \citenamefont {Maitland}, \citenamefont
  {Marino}, \citenamefont {Prain}, \citenamefont {Starshynov}, \citenamefont
  {Westerberg}, \citenamefont {Wright},\ and\ \citenamefont
  {Faccio}}]{braidotti2022measurement}%
  \BibitemOpen
  \bibfield  {author} {\bibinfo {author} {\bibfnamefont {M.~C.}\ \bibnamefont
  {Braidotti}}, \bibinfo {author} {\bibfnamefont {R.}~\bibnamefont {Prizia}},
  \bibinfo {author} {\bibfnamefont {C.}~\bibnamefont {Maitland}}, \bibinfo
  {author} {\bibfnamefont {F.}~\bibnamefont {Marino}}, \bibinfo {author}
  {\bibfnamefont {A.}~\bibnamefont {Prain}}, \bibinfo {author} {\bibfnamefont
  {I.}~\bibnamefont {Starshynov}}, \bibinfo {author} {\bibfnamefont
  {N.}~\bibnamefont {Westerberg}}, \bibinfo {author} {\bibfnamefont {E.~M.}\
  \bibnamefont {Wright}},\ and\ \bibinfo {author} {\bibfnamefont
  {D.}~\bibnamefont {Faccio}},\ }\bibfield  {title} {\bibinfo {title}
  {{Measurement of Penrose Superradiance in a Photon Superfluid}},\ }\href
  {https://doi.org/10.1103/PhysRevLett.128.013901} {\bibfield  {journal}
  {\bibinfo  {journal} {Phys. Rev. Lett.}\ }\textbf {\bibinfo {volume} {128}},\
  \bibinfo {pages} {013901} (\bibinfo {year} {2022})}\BibitemShut {NoStop}%
\bibitem [{\citenamefont {Richartz}\ \emph {et~al.}(2015)\citenamefont
  {Richartz}, \citenamefont {Prain}, \citenamefont {Liberati},\ and\
  \citenamefont {Weinfurtner}}]{PhysRevD.91.124018}%
  \BibitemOpen
  \bibfield  {author} {\bibinfo {author} {\bibfnamefont {M.}~\bibnamefont
  {Richartz}}, \bibinfo {author} {\bibfnamefont {A.}~\bibnamefont {Prain}},
  \bibinfo {author} {\bibfnamefont {S.}~\bibnamefont {Liberati}},\ and\
  \bibinfo {author} {\bibfnamefont {S.}~\bibnamefont {Weinfurtner}},\
  }\bibfield  {title} {\bibinfo {title} {Rotating black holes in a draining
  bathtub: Superradiant scattering of gravity waves},\ }\href
  {https://doi.org/10.1103/PhysRevD.91.124018} {\bibfield  {journal} {\bibinfo
  {journal} {Phys. Rev. D}\ }\textbf {\bibinfo {volume} {91}},\ \bibinfo
  {pages} {124018} (\bibinfo {year} {2015})}\BibitemShut {NoStop}%
\bibitem [{\citenamefont {Garay}\ \emph {et~al.}(2000)\citenamefont {Garay},
  \citenamefont {Anglin}, \citenamefont {Cirac},\ and\ \citenamefont
  {Zoller}}]{PhysRevLett85.4643}%
  \BibitemOpen
  \bibfield  {author} {\bibinfo {author} {\bibfnamefont {L.~J.}\ \bibnamefont
  {Garay}}, \bibinfo {author} {\bibfnamefont {J.~R.}\ \bibnamefont {Anglin}},
  \bibinfo {author} {\bibfnamefont {J.~I.}\ \bibnamefont {Cirac}},\ and\
  \bibinfo {author} {\bibfnamefont {P.}~\bibnamefont {Zoller}},\ }\bibfield
  {title} {\bibinfo {title} {{Sonic Analog of Gravitational Black Holes in
  Bose-Einstein Condensates}},\ }\href
  {https://doi.org/10.1103/PhysRevLett.85.4643} {\bibfield  {journal} {\bibinfo
   {journal} {Phys. Rev. Lett.}\ }\textbf {\bibinfo {volume} {85}},\ \bibinfo
  {pages} {4643} (\bibinfo {year} {2000})}\BibitemShut {NoStop}%
\bibitem [{\citenamefont {Barcel{\'{o}}}\ \emph
  {et~al.}(2001{\natexlab{b}})\citenamefont {Barcel{\'{o}}}, \citenamefont
  {Liberati},\ and\ \citenamefont {Visser}}]{barcelo2001analogue}%
  \BibitemOpen
  \bibfield  {author} {\bibinfo {author} {\bibfnamefont {C.}~\bibnamefont
  {Barcel{\'{o}}}}, \bibinfo {author} {\bibfnamefont {S.}~\bibnamefont
  {Liberati}},\ and\ \bibinfo {author} {\bibfnamefont {M.}~\bibnamefont
  {Visser}},\ }\bibfield  {title} {\bibinfo {title} {{Analogue gravity from
  Bose-Einstein condensates}},\ }\href
  {https://doi.org/10.1088/0264-9381/18/6/312} {\bibfield  {journal} {\bibinfo
  {journal} {Classical and Quantum Gravity}\ }\textbf {\bibinfo {volume}
  {18}},\ \bibinfo {pages} {1137} (\bibinfo {year}
  {2001}{\natexlab{b}})}\BibitemShut {NoStop}%
\bibitem [{\citenamefont {Carusotto}\ \emph {et~al.}(2008)\citenamefont
  {Carusotto}, \citenamefont {Fagnocchi}, \citenamefont {Recati}, \citenamefont
  {Balbinot},\ and\ \citenamefont {Fabbri}}]{Carusotto_2008}%
  \BibitemOpen
  \bibfield  {author} {\bibinfo {author} {\bibfnamefont {I.}~\bibnamefont
  {Carusotto}}, \bibinfo {author} {\bibfnamefont {S.}~\bibnamefont
  {Fagnocchi}}, \bibinfo {author} {\bibfnamefont {A.}~\bibnamefont {Recati}},
  \bibinfo {author} {\bibfnamefont {R.}~\bibnamefont {Balbinot}},\ and\
  \bibinfo {author} {\bibfnamefont {A.}~\bibnamefont {Fabbri}},\ }\bibfield
  {title} {\bibinfo {title} {{Numerical observation of Hawking radiation from
  acoustic black holes in atomic Bose{\textendash}Einstein condensates}},\
  }\href {https://doi.org/10.1088/1367-2630/10/10/103001} {\bibfield  {journal}
  {\bibinfo  {journal} {New Journal of Physics}\ }\textbf {\bibinfo {volume}
  {10}},\ \bibinfo {pages} {103001} (\bibinfo {year} {2008})}\BibitemShut
  {NoStop}%
\bibitem [{\citenamefont {Macher}\ and\ \citenamefont
  {Parentani}(2009)}]{Macher}%
  \BibitemOpen
  \bibfield  {author} {\bibinfo {author} {\bibfnamefont {J.}~\bibnamefont
  {Macher}}\ and\ \bibinfo {author} {\bibfnamefont {R.}~\bibnamefont
  {Parentani}},\ }\bibfield  {title} {\bibinfo {title} {{Black-hole radiation
  in Bose-Einstein condensates}},\ }\href
  {https://doi.org/10.1103/PhysRevA.80.043601} {\bibfield  {journal} {\bibinfo
  {journal} {Phys. Rev. A}\ }\textbf {\bibinfo {volume} {80}},\ \bibinfo
  {pages} {043601} (\bibinfo {year} {2009})}\BibitemShut {NoStop}%
\bibitem [{\citenamefont {Lahav}\ \emph {et~al.}(2010)\citenamefont {Lahav},
  \citenamefont {Itah}, \citenamefont {Blumkin}, \citenamefont {Gordon},
  \citenamefont {Rinott}, \citenamefont {Zayats},\ and\ \citenamefont
  {Steinhauer}}]{Lahav}%
  \BibitemOpen
  \bibfield  {author} {\bibinfo {author} {\bibfnamefont {O.}~\bibnamefont
  {Lahav}}, \bibinfo {author} {\bibfnamefont {A.}~\bibnamefont {Itah}},
  \bibinfo {author} {\bibfnamefont {A.}~\bibnamefont {Blumkin}}, \bibinfo
  {author} {\bibfnamefont {C.}~\bibnamefont {Gordon}}, \bibinfo {author}
  {\bibfnamefont {S.}~\bibnamefont {Rinott}}, \bibinfo {author} {\bibfnamefont
  {A.}~\bibnamefont {Zayats}},\ and\ \bibinfo {author} {\bibfnamefont
  {J.}~\bibnamefont {Steinhauer}},\ }\bibfield  {title} {\bibinfo {title}
  {{Realization of a Sonic Black Hole Analog in a Bose-Einstein Condensate}},\
  }\href {https://doi.org/10.1103/PhysRevLett.105.240401} {\bibfield  {journal}
  {\bibinfo  {journal} {Phys. Rev. Lett.}\ }\textbf {\bibinfo {volume} {105}},\
  \bibinfo {pages} {240401} (\bibinfo {year} {2010})}\BibitemShut {NoStop}%
\bibitem [{\citenamefont {Steinhauer}(2016)}]{Steinhauer16}%
  \BibitemOpen
  \bibfield  {author} {\bibinfo {author} {\bibfnamefont {J.}~\bibnamefont
  {Steinhauer}},\ }\bibfield  {title} {\bibinfo {title} {{Observation of
  quantum Hawking radiation and its entanglement in an analogue black hole}},\
  }\href {http://dx.doi.org/10.1038/nphys3863} {\bibfield  {journal} {\bibinfo
  {journal} {Nat. Phys.}\ }\textbf {\bibinfo {volume} {12}},\ \bibinfo {pages}
  {959} (\bibinfo {year} {2016})}\BibitemShut {NoStop}%
\bibitem [{\citenamefont {Mu{\~n}oz~de Nova}\ \emph {et~al.}(2019)\citenamefont
  {Mu{\~n}oz~de Nova}, \citenamefont {Golubkov}, \citenamefont {Kolobov},\ and\
  \citenamefont {Steinhauer}}]{Munoz}%
  \BibitemOpen
  \bibfield  {author} {\bibinfo {author} {\bibfnamefont {J.~R.}\ \bibnamefont
  {Mu{\~n}oz~de Nova}}, \bibinfo {author} {\bibfnamefont {K.}~\bibnamefont
  {Golubkov}}, \bibinfo {author} {\bibfnamefont {V.~I.}\ \bibnamefont
  {Kolobov}},\ and\ \bibinfo {author} {\bibfnamefont {J.}~\bibnamefont
  {Steinhauer}},\ }\bibfield  {title} {\bibinfo {title} {{Observation of
  thermal Hawking radiation and its temperature in an analogue black hole}},\
  }\href {https://doi.org/10.1038/s41586-019-1241-0} {\bibfield  {journal}
  {\bibinfo  {journal} {Nature}\ }\textbf {\bibinfo {volume} {569}},\ \bibinfo
  {pages} {688} (\bibinfo {year} {2019})}\BibitemShut {NoStop}%
\bibitem [{\citenamefont {Ch\"a}\ and\ \citenamefont
  {Fischer}(2017)}]{PhysRevLett118.130404}%
  \BibitemOpen
  \bibfield  {author} {\bibinfo {author} {\bibfnamefont {S.-Y.}\ \bibnamefont
  {Ch\"a}}\ and\ \bibinfo {author} {\bibfnamefont {U.~R.}\ \bibnamefont
  {Fischer}},\ }\bibfield  {title} {\bibinfo {title} {{Probing the Scale
  Invariance of the Inflationary Power Spectrum in Expanding
  Quasi-Two-Dimensional Dipolar Condensates}},\ }\href
  {https://doi.org/10.1103/PhysRevLett.118.130404} {\bibfield  {journal}
  {\bibinfo  {journal} {Phys. Rev. Lett.}\ }\textbf {\bibinfo {volume} {118}},\
  \bibinfo {pages} {130404} (\bibinfo {year} {2017})}\BibitemShut {NoStop}%
\bibitem [{\citenamefont {Eckel}\ \emph {et~al.}(2018)\citenamefont {Eckel},
  \citenamefont {Kumar}, \citenamefont {Jacobson}, \citenamefont {Spielman},\
  and\ \citenamefont {Campbell}}]{Eckel}%
  \BibitemOpen
  \bibfield  {author} {\bibinfo {author} {\bibfnamefont {S.}~\bibnamefont
  {Eckel}}, \bibinfo {author} {\bibfnamefont {A.}~\bibnamefont {Kumar}},
  \bibinfo {author} {\bibfnamefont {T.}~\bibnamefont {Jacobson}}, \bibinfo
  {author} {\bibfnamefont {I.~B.}\ \bibnamefont {Spielman}},\ and\ \bibinfo
  {author} {\bibfnamefont {G.~K.}\ \bibnamefont {Campbell}},\ }\bibfield
  {title} {\bibinfo {title} {{A Rapidly Expanding Bose-Einstein Condensate: An
  Expanding Universe in the Lab}},\ }\href
  {https://doi.org/10.1103/PhysRevX.8.021021} {\bibfield  {journal} {\bibinfo
  {journal} {Phys. Rev. X}\ }\textbf {\bibinfo {volume} {8}},\ \bibinfo {pages}
  {021021} (\bibinfo {year} {2018})}\BibitemShut {NoStop}%
\bibitem [{\citenamefont {Eckel}\ and\ \citenamefont
  {Jacobson}(2021)}]{Eckel2}%
  \BibitemOpen
  \bibfield  {author} {\bibinfo {author} {\bibfnamefont {S.}~\bibnamefont
  {Eckel}}\ and\ \bibinfo {author} {\bibfnamefont {T.}~\bibnamefont
  {Jacobson}},\ }\bibfield  {title} {\bibinfo {title} {{Phonon redshift and
  Hubble friction in an expanding BEC}},\ }\href
  {https://doi.org/10.21468/SciPostPhys.10.3.064} {\bibfield  {journal}
  {\bibinfo  {journal} {SciPost Phys.}\ }\textbf {\bibinfo {volume} {10}},\
  \bibinfo {pages} {64} (\bibinfo {year} {2021})}\BibitemShut {NoStop}%
\bibitem [{\citenamefont {Banik}\ \emph {et~al.}(2022)\citenamefont {Banik},
  \citenamefont {Galan}, \citenamefont {Sosa-Martinez}, \citenamefont
  {Anderson}, \citenamefont {Eckel}, \citenamefont {Spielman},\ and\
  \citenamefont {Campbell}}]{Banik}%
  \BibitemOpen
  \bibfield  {author} {\bibinfo {author} {\bibfnamefont {S.}~\bibnamefont
  {Banik}}, \bibinfo {author} {\bibfnamefont {M.~G.}\ \bibnamefont {Galan}},
  \bibinfo {author} {\bibfnamefont {H.}~\bibnamefont {Sosa-Martinez}}, \bibinfo
  {author} {\bibfnamefont {M.~J.}\ \bibnamefont {Anderson}}, \bibinfo {author}
  {\bibfnamefont {S.}~\bibnamefont {Eckel}}, \bibinfo {author} {\bibfnamefont
  {I.~B.}\ \bibnamefont {Spielman}},\ and\ \bibinfo {author} {\bibfnamefont
  {G.~K.}\ \bibnamefont {Campbell}},\ }\bibfield  {title} {\bibinfo {title}
  {{Accurate Determination of Hubble Attenuation and Amplification in Expanding
  and Contracting Cold-Atom Universes}},\ }\href
  {https://doi.org/10.1103/PhysRevLett.128.090401} {\bibfield  {journal}
  {\bibinfo  {journal} {Phys. Rev. Lett.}\ }\textbf {\bibinfo {volume} {128}},\
  \bibinfo {pages} {090401} (\bibinfo {year} {2022})}\BibitemShut {NoStop}%
\bibitem [{\citenamefont {Fischer}\ and\ \citenamefont
  {Sch\"utzhold}(2004)}]{Schuetzhold}%
  \BibitemOpen
  \bibfield  {author} {\bibinfo {author} {\bibfnamefont {U.~R.}\ \bibnamefont
  {Fischer}}\ and\ \bibinfo {author} {\bibfnamefont {R.}~\bibnamefont
  {Sch\"utzhold}},\ }\bibfield  {title} {\bibinfo {title} {{Quantum simulation
  of cosmic inflation in two-component Bose-Einstein condensates}},\ }\href
  {https://doi.org/10.1103/PhysRevA.70.063615} {\bibfield  {journal} {\bibinfo
  {journal} {Phys. Rev. A}\ }\textbf {\bibinfo {volume} {70}},\ \bibinfo
  {pages} {063615} (\bibinfo {year} {2004})}\BibitemShut {NoStop}%
\bibitem [{\citenamefont {Barcel\'o}\ \emph {et~al.}(2003)\citenamefont
  {Barcel\'o}, \citenamefont {Liberati},\ and\ \citenamefont
  {Visser}}]{BLV2003PRA}%
  \BibitemOpen
  \bibfield  {author} {\bibinfo {author} {\bibfnamefont {C.}~\bibnamefont
  {Barcel\'o}}, \bibinfo {author} {\bibfnamefont {S.}~\bibnamefont
  {Liberati}},\ and\ \bibinfo {author} {\bibfnamefont {M.}~\bibnamefont
  {Visser}},\ }\bibfield  {title} {\bibinfo {title} {{Probing semiclassical
  analog gravity in Bose-Einstein condensates with widely tunable
  interactions}},\ }\href {https://doi.org/10.1103/PhysRevA.68.053613}
  {\bibfield  {journal} {\bibinfo  {journal} {Phys. Rev. A}\ }\textbf {\bibinfo
  {volume} {68}},\ \bibinfo {pages} {053613} (\bibinfo {year}
  {2003})}\BibitemShut {NoStop}%
\bibitem [{\citenamefont {Fedichev}\ and\ \citenamefont {Fischer}(2004)}]{CPP}%
  \BibitemOpen
  \bibfield  {author} {\bibinfo {author} {\bibfnamefont {P.~O.}\ \bibnamefont
  {Fedichev}}\ and\ \bibinfo {author} {\bibfnamefont {U.~R.}\ \bibnamefont
  {Fischer}},\ }\bibfield  {title} {\bibinfo {title} {{``Cosmological''
  quasiparticle production in harmonically trapped superfluid gases}},\ }\href
  {https://doi.org/10.1103/PhysRevA.69.033602} {\bibfield  {journal} {\bibinfo
  {journal} {Phys. Rev. A}\ }\textbf {\bibinfo {volume} {69}},\ \bibinfo
  {pages} {033602} (\bibinfo {year} {2004})}\BibitemShut {NoStop}%
\bibitem [{\citenamefont {Robertson}\ \emph {et~al.}(2017)\citenamefont
  {Robertson}, \citenamefont {Michel},\ and\ \citenamefont
  {Parentani}}]{Robertson}%
  \BibitemOpen
  \bibfield  {author} {\bibinfo {author} {\bibfnamefont {S.}~\bibnamefont
  {Robertson}}, \bibinfo {author} {\bibfnamefont {F.}~\bibnamefont {Michel}},\
  and\ \bibinfo {author} {\bibfnamefont {R.}~\bibnamefont {Parentani}},\
  }\bibfield  {title} {\bibinfo {title} {{Assessing degrees of entanglement of
  phonon states in atomic Bose gases through the measurement of commuting
  observables}},\ }\href {https://doi.org/10.1103/PhysRevD.96.045012}
  {\bibfield  {journal} {\bibinfo  {journal} {Phys. Rev. D}\ }\textbf {\bibinfo
  {volume} {96}},\ \bibinfo {pages} {045012} (\bibinfo {year}
  {2017})}\BibitemShut {NoStop}%
\bibitem [{\citenamefont {Gooding}\ \emph {et~al.}(2020)\citenamefont
  {Gooding}, \citenamefont {Biermann}, \citenamefont {Erne}, \citenamefont
  {Louko}, \citenamefont {Unruh}, \citenamefont {Schmiedmayer},\ and\
  \citenamefont {Weinfurtner}}]{Gooding}%
  \BibitemOpen
  \bibfield  {author} {\bibinfo {author} {\bibfnamefont {C.}~\bibnamefont
  {Gooding}}, \bibinfo {author} {\bibfnamefont {S.}~\bibnamefont {Biermann}},
  \bibinfo {author} {\bibfnamefont {S.}~\bibnamefont {Erne}}, \bibinfo {author}
  {\bibfnamefont {J.}~\bibnamefont {Louko}}, \bibinfo {author} {\bibfnamefont
  {W.~G.}\ \bibnamefont {Unruh}}, \bibinfo {author} {\bibfnamefont
  {J.}~\bibnamefont {Schmiedmayer}},\ and\ \bibinfo {author} {\bibfnamefont
  {S.}~\bibnamefont {Weinfurtner}},\ }\bibfield  {title} {\bibinfo {title}
  {{Interferometric Unruh Detectors for Bose-Einstein Condensates}},\ }\href
  {https://doi.org/10.1103/PhysRevLett.125.213603} {\bibfield  {journal}
  {\bibinfo  {journal} {Phys. Rev. Lett.}\ }\textbf {\bibinfo {volume} {125}},\
  \bibinfo {pages} {213603} (\bibinfo {year} {2020})}\BibitemShut {NoStop}%
\bibitem [{\citenamefont {Finazzi}\ and\ \citenamefont
  {Parentani}(2010)}]{Finazzi}%
  \BibitemOpen
  \bibfield  {author} {\bibinfo {author} {\bibfnamefont {S.}~\bibnamefont
  {Finazzi}}\ and\ \bibinfo {author} {\bibfnamefont {R.}~\bibnamefont
  {Parentani}},\ }\bibfield  {title} {\bibinfo {title} {{Black hole lasers in
  Bose{\textendash}Einstein condensates}},\ }\href
  {https://doi.org/10.1088/1367-2630/12/9/095015} {\bibfield  {journal}
  {\bibinfo  {journal} {New Journal of Physics}\ }\textbf {\bibinfo {volume}
  {12}},\ \bibinfo {pages} {095015} (\bibinfo {year} {2010})}\BibitemShut
  {NoStop}%
\bibitem [{\citenamefont {Tian}\ \emph {et~al.}(2018)\citenamefont {Tian},
  \citenamefont {Ch\"a},\ and\ \citenamefont {Fischer}}]{Tian}%
  \BibitemOpen
  \bibfield  {author} {\bibinfo {author} {\bibfnamefont {Z.}~\bibnamefont
  {Tian}}, \bibinfo {author} {\bibfnamefont {S.-Y.}\ \bibnamefont {Ch\"a}},\
  and\ \bibinfo {author} {\bibfnamefont {U.~R.}\ \bibnamefont {Fischer}},\
  }\bibfield  {title} {\bibinfo {title} {{Roton entanglement in quenched
  dipolar Bose-Einstein condensates}},\ }\href
  {https://doi.org/10.1103/PhysRevA.97.063611} {\bibfield  {journal} {\bibinfo
  {journal} {Phys. Rev. A}\ }\textbf {\bibinfo {volume} {97}},\ \bibinfo
  {pages} {063611} (\bibinfo {year} {2018})}\BibitemShut {NoStop}%
\bibitem [{\citenamefont {Fedichev}\ and\ \citenamefont
  {Fischer}(2003)}]{Fedichev}%
  \BibitemOpen
  \bibfield  {author} {\bibinfo {author} {\bibfnamefont {P.~O.}\ \bibnamefont
  {Fedichev}}\ and\ \bibinfo {author} {\bibfnamefont {U.~R.}\ \bibnamefont
  {Fischer}},\ }\bibfield  {title} {\bibinfo {title} {{Gibbons-Hawking Effect
  in the Sonic de Sitter Space-Time of an Expanding Bose-Einstein-Condensed
  Gas}},\ }\href {https://doi.org/10.1103/PhysRevLett.91.240407} {\bibfield
  {journal} {\bibinfo  {journal} {Phys. Rev. Lett.}\ }\textbf {\bibinfo
  {volume} {91}},\ \bibinfo {pages} {240407} (\bibinfo {year}
  {2003})}\BibitemShut {NoStop}%
\bibitem [{\citenamefont {Retzker}\ \emph {et~al.}(2008)\citenamefont
  {Retzker}, \citenamefont {Cirac}, \citenamefont {Plenio},\ and\ \citenamefont
  {Reznik}}]{Reznik}%
  \BibitemOpen
  \bibfield  {author} {\bibinfo {author} {\bibfnamefont {A.}~\bibnamefont
  {Retzker}}, \bibinfo {author} {\bibfnamefont {J.~I.}\ \bibnamefont {Cirac}},
  \bibinfo {author} {\bibfnamefont {M.~B.}\ \bibnamefont {Plenio}},\ and\
  \bibinfo {author} {\bibfnamefont {B.}~\bibnamefont {Reznik}},\ }\bibfield
  {title} {\bibinfo {title} {{Methods for Detecting Acceleration Radiation in a
  Bose-Einstein Condensate}},\ }\href
  {https://doi.org/10.1103/PhysRevLett.101.110402} {\bibfield  {journal}
  {\bibinfo  {journal} {Phys. Rev. Lett.}\ }\textbf {\bibinfo {volume} {101}},\
  \bibinfo {pages} {110402} (\bibinfo {year} {2008})}\BibitemShut {NoStop}%
\bibitem [{\citenamefont {Hartley}\ \emph {et~al.}(2018)\citenamefont
  {Hartley}, \citenamefont {Bravo}, \citenamefont {R\"atzel}, \citenamefont
  {Howl},\ and\ \citenamefont {Fuentes}}]{hartley2018analogue}%
  \BibitemOpen
  \bibfield  {author} {\bibinfo {author} {\bibfnamefont {D.}~\bibnamefont
  {Hartley}}, \bibinfo {author} {\bibfnamefont {T.}~\bibnamefont {Bravo}},
  \bibinfo {author} {\bibfnamefont {D.}~\bibnamefont {R\"atzel}}, \bibinfo
  {author} {\bibfnamefont {R.}~\bibnamefont {Howl}},\ and\ \bibinfo {author}
  {\bibfnamefont {I.}~\bibnamefont {Fuentes}},\ }\bibfield  {title} {\bibinfo
  {title} {{Analogue simulation of gravitational waves in a $3+1$-dimensional
  Bose-Einstein condensate}},\ }\href
  {https://doi.org/10.1103/PhysRevD.98.025011} {\bibfield  {journal} {\bibinfo
  {journal} {Phys. Rev. D}\ }\textbf {\bibinfo {volume} {98}},\ \bibinfo
  {pages} {025011} (\bibinfo {year} {2018})}\BibitemShut {NoStop}%
\bibitem [{\citenamefont {Datta}\ and\ \citenamefont
  {Fischer}(2022{\natexlab{a}})}]{PhysRevD.105.022003}%
  \BibitemOpen
  \bibfield  {author} {\bibinfo {author} {\bibfnamefont {S.}~\bibnamefont
  {Datta}}\ and\ \bibinfo {author} {\bibfnamefont {U.~R.}\ \bibnamefont
  {Fischer}},\ }\bibfield  {title} {\bibinfo {title} {Inherent nonlinearity of
  fluid motion and acoustic gravitational wave memory},\ }\href
  {https://doi.org/10.1103/PhysRevD.105.022003} {\bibfield  {journal} {\bibinfo
   {journal} {Phys. Rev. D}\ }\textbf {\bibinfo {volume} {105}},\ \bibinfo
  {pages} {022003} (\bibinfo {year} {2022}{\natexlab{a}})}\BibitemShut
  {NoStop}%
\bibitem [{\citenamefont {Unruh}()}]{Unruh1994}%
  \BibitemOpen
  \bibfield  {author} {\bibinfo {author} {\bibfnamefont {W.~G.}\ \bibnamefont
  {Unruh}},\ }\bibfield  {title} {\bibinfo {title} {Dumb holes and the effects
  of high frequencies on black hole evaporation},\ }\href@noop {} {\ }\Eprint
  {https://arxiv.org/abs/gr-qc/9409008} {arXiv:gr-qc/9409008 [gr-qc]}
  \BibitemShut {NoStop}%
\bibitem [{\citenamefont {Datta}\ and\ \citenamefont
  {Fischer}(2022{\natexlab{b}})}]{Datta_2022}%
  \BibitemOpen
  \bibfield  {author} {\bibinfo {author} {\bibfnamefont {S.}~\bibnamefont
  {Datta}}\ and\ \bibinfo {author} {\bibfnamefont {U.~R.}\ \bibnamefont
  {Fischer}},\ }\bibfield  {title} {\bibinfo {title} {Analogue gravitational
  field from nonlinear fluid dynamics},\ }\href
  {https://doi.org/10.1088/1361-6382/ac4828} {\bibfield  {journal} {\bibinfo
  {journal} {Classical and Quantum Gravity}\ }\textbf {\bibinfo {volume}
  {39}},\ \bibinfo {pages} {075018} (\bibinfo {year}
  {2022}{\natexlab{b}})}\BibitemShut {NoStop}%
\bibitem [{\citenamefont {Novello}\ \emph {et~al.}(2013)\citenamefont
  {Novello}, \citenamefont {Bittencourt}, \citenamefont {Moschella},
  \citenamefont {Goulart}, \citenamefont {Salim},\ and\ \citenamefont
  {Toniato}}]{Novello_2013}%
  \BibitemOpen
  \bibfield  {author} {\bibinfo {author} {\bibfnamefont {M.}~\bibnamefont
  {Novello}}, \bibinfo {author} {\bibfnamefont {E.}~\bibnamefont
  {Bittencourt}}, \bibinfo {author} {\bibfnamefont {U.}~\bibnamefont
  {Moschella}}, \bibinfo {author} {\bibfnamefont {E.}~\bibnamefont {Goulart}},
  \bibinfo {author} {\bibfnamefont {J.~M.}\ \bibnamefont {Salim}},\ and\
  \bibinfo {author} {\bibfnamefont {J.~D.}\ \bibnamefont {Toniato}},\
  }\bibfield  {title} {\bibinfo {title} {Geometric scalar theory of gravity},\
  }\href {https://doi.org/10.1088/1475-7516/2013/06/014} {\bibfield  {journal}
  {\bibinfo  {journal} {Journal of Cosmology and Astroparticle Physics}\
  }\textbf {\bibinfo {volume} {2013}}\bibinfo  {number} { (06)},\ \bibinfo
  {pages} {014}}\BibitemShut {NoStop}%
\bibitem [{\citenamefont {Landau}\ and\ \citenamefont
  {Lifshitz}(1987)}]{Landau1987Fluid}%
  \BibitemOpen
\bibfield  {number} {  }\bibfield  {author} {\bibinfo {author} {\bibfnamefont
  {L.~D.}\ \bibnamefont {Landau}}\ and\ \bibinfo {author} {\bibfnamefont
  {E.~M.}\ \bibnamefont {Lifshitz}},\ }\href
  {http://www.worldcat.org/isbn/0750627670} {\emph {\bibinfo {title} {Fluid
  Mechanics, Second Edition: Volume 6 (Course of Theoretical Physics)}}},\
  \bibinfo {edition} {2nd}\ ed.\ (\bibinfo  {publisher}
  {Butterworth-Heinemann},\ \bibinfo {year} {1987})\BibitemShut {NoStop}%
\bibitem [{\citenamefont {Dalfovo}\ \emph {et~al.}(1999)\citenamefont
  {Dalfovo}, \citenamefont {Giorgini}, \citenamefont {Pitaevski\v\i},\ and\
  \citenamefont {Stringari}}]{RevModPhys.71.463}%
  \BibitemOpen
  \bibfield  {author} {\bibinfo {author} {\bibfnamefont {F.}~\bibnamefont
  {Dalfovo}}, \bibinfo {author} {\bibfnamefont {S.}~\bibnamefont {Giorgini}},
  \bibinfo {author} {\bibfnamefont {L.~P.}\ \bibnamefont {Pitaevski\v\i}},\
  and\ \bibinfo {author} {\bibfnamefont {S.}~\bibnamefont {Stringari}},\
  }\bibfield  {title} {\bibinfo {title} {{Theory of Bose-Einstein condensation
  in trapped gases}},\ }\href {https://doi.org/10.1103/RevModPhys.71.463}
  {\bibfield  {journal} {\bibinfo  {journal} {Rev. Mod. Phys.}\ }\textbf
  {\bibinfo {volume} {71}},\ \bibinfo {pages} {463} (\bibinfo {year}
  {1999})}\BibitemShut {NoStop}%
\bibitem [{\citenamefont {Damski}(2004)}]{Damski2004}%
  \BibitemOpen
  \bibfield  {author} {\bibinfo {author} {\bibfnamefont {B.}~\bibnamefont
  {Damski}},\ }\bibfield  {title} {\bibinfo {title} {{Formation of shock waves
  in a Bose-Einstein condensate}},\ }\href
  {https://doi.org/10.1103/PhysRevA.69.043610} {\bibfield  {journal} {\bibinfo
  {journal} {Phys. Rev. A}\ }\textbf {\bibinfo {volume} {69}},\ \bibinfo
  {pages} {043610} (\bibinfo {year} {2004})}\BibitemShut {NoStop}%
\bibitem [{\citenamefont {Visser}(2007)}]{Visser:2007nx}%
  \BibitemOpen
  \bibfield  {author} {\bibinfo {author} {\bibfnamefont {M.}~\bibnamefont
  {Visser}},\ }\bibfield  {title} {\bibinfo {title} {{{Emergent rainbow
  spacetimes: Two pedagogical examples}}},\ }in\ \href@noop {} {\emph {\bibinfo
  {booktitle} {{Time and Matter 2007}}}}\ (\bibinfo  {publisher} {University of
  Nova Gorica Press},\ \bibinfo {year} {2007})\ pp.\ \bibinfo {pages}
  {191--205},\ \Eprint {https://arxiv.org/abs/0712.0810} {arXiv:0712.0810
  [gr-qc]} \BibitemShut {NoStop}%
\bibitem [{\citenamefont {Weinfurtner}\ \emph {et~al.}(2009)\citenamefont
  {Weinfurtner}, \citenamefont {Jain}, \citenamefont {Visser},\ and\
  \citenamefont {Gardiner}}]{Weinfurtner_2009}%
  \BibitemOpen
  \bibfield  {author} {\bibinfo {author} {\bibfnamefont {S.}~\bibnamefont
  {Weinfurtner}}, \bibinfo {author} {\bibfnamefont {P.}~\bibnamefont {Jain}},
  \bibinfo {author} {\bibfnamefont {M.}~\bibnamefont {Visser}},\ and\ \bibinfo
  {author} {\bibfnamefont {C.~W.}\ \bibnamefont {Gardiner}},\ }\bibfield
  {title} {\bibinfo {title} {Cosmological particle production in emergent
  rainbow spacetimes},\ }\href {https://doi.org/10.1088/0264-9381/26/6/065012}
  {\bibfield  {journal} {\bibinfo  {journal} {Classical and Quantum Gravity}\
  }\textbf {\bibinfo {volume} {26}},\ \bibinfo {pages} {065012} (\bibinfo
  {year} {2009})}\BibitemShut {NoStop}%
\bibitem [{\citenamefont {Meppelink}\ \emph {et~al.}(2009)\citenamefont
  {Meppelink}, \citenamefont {Koller}, \citenamefont {Vogels}, \citenamefont
  {van~der Straten}, \citenamefont {van Ooijen}, \citenamefont {Heckenberg},
  \citenamefont {Rubinsztein-Dunlop}, \citenamefont {Haine},\ and\
  \citenamefont {Davis}}]{meppelink2009observation}%
  \BibitemOpen
  \bibfield  {author} {\bibinfo {author} {\bibfnamefont {R.}~\bibnamefont
  {Meppelink}}, \bibinfo {author} {\bibfnamefont {S.~B.}\ \bibnamefont
  {Koller}}, \bibinfo {author} {\bibfnamefont {J.~M.}\ \bibnamefont {Vogels}},
  \bibinfo {author} {\bibfnamefont {P.}~\bibnamefont {van~der Straten}},
  \bibinfo {author} {\bibfnamefont {E.~D.}\ \bibnamefont {van Ooijen}},
  \bibinfo {author} {\bibfnamefont {N.~R.}\ \bibnamefont {Heckenberg}},
  \bibinfo {author} {\bibfnamefont {H.}~\bibnamefont {Rubinsztein-Dunlop}},
  \bibinfo {author} {\bibfnamefont {S.~A.}\ \bibnamefont {Haine}},\ and\
  \bibinfo {author} {\bibfnamefont {M.~J.}\ \bibnamefont {Davis}},\ }\bibfield
  {title} {\bibinfo {title} {{Observation of shock waves in a large
  Bose-Einstein condensate}},\ }\href
  {https://doi.org/10.1103/PhysRevA.80.043606} {\bibfield  {journal} {\bibinfo
  {journal} {Phys. Rev. A}\ }\textbf {\bibinfo {volume} {80}},\ \bibinfo
  {pages} {043606} (\bibinfo {year} {2009})}\BibitemShut {NoStop}%
\bibitem [{\citenamefont {Marino}\ \emph {et~al.}(2015)\citenamefont {Marino},
  \citenamefont {Maitland}, \citenamefont {Vocke}, \citenamefont {Ortolan},\
  and\ \citenamefont {Faccio}}]{marino2016emergent}%
  \BibitemOpen
  \bibfield  {author} {\bibinfo {author} {\bibfnamefont {F.}~\bibnamefont
  {Marino}}, \bibinfo {author} {\bibfnamefont {C.}~\bibnamefont {Maitland}},
  \bibinfo {author} {\bibfnamefont {D.}~\bibnamefont {Vocke}}, \bibinfo
  {author} {\bibfnamefont {A.}~\bibnamefont {Ortolan}},\ and\ \bibinfo {author}
  {\bibfnamefont {D.}~\bibnamefont {Faccio}},\ }\bibfield  {title} {\bibinfo
  {title} {Emergent geometries and nonlinear-wave dynamics in photon fluids},\
  }\href {https://doi.org/10.1038/srep23282} {\bibfield  {journal} {\bibinfo
  {journal} {Scientific Reports}\ }\textbf {\bibinfo {volume} {6}} (\bibinfo
  {year} {2015})}\BibitemShut {NoStop}%
\bibitem [{\citenamefont {Shen}(2016)}]{shen2016introduction}%
  \BibitemOpen
  \bibfield  {author} {\bibinfo {author} {\bibfnamefont {W.}~\bibnamefont
  {Shen}},\ }\href {https://books.google.co.kr/books?id=AO2djgEACAAJ} {\emph
  {\bibinfo {title} {An Introduction to Numerical Computation}}}\ (\bibinfo
  {publisher} {World Scientific},\ \bibinfo {year} {2016})\BibitemShut
  {NoStop}%
\bibitem [{\citenamefont {Riemann}(1860)}]{Riemann1860}%
  \BibitemOpen
  \bibfield  {author} {\bibinfo {author} {\bibfnamefont {B.}~\bibnamefont
  {Riemann}},\ }\bibfield  {title} {\bibinfo {title} {{Ueber die Fortpflanzung
  ebener Luftwellen von endlicher Schwingungsweite}},\ }\href
  {http://eudml.org/doc/135717} {\bibfield  {journal} {\bibinfo  {journal}
  {Abhandlungen der K\"oniglichen Gesellschaft der Wissenschaften in
  G\"ottingen}\ }\textbf {\bibinfo {volume} {8}},\ \bibinfo {pages} {43}
  (\bibinfo {year} {1860})}\BibitemShut {NoStop}%
\bibitem [{\citenamefont {Kersal{\'e}}(2004)}]{PDES}%
  \BibitemOpen
  \bibfield  {author} {\bibinfo {author} {\bibfnamefont {E.}~\bibnamefont
  {Kersal{\'e}}},\ }\bibfield  {title} {\bibinfo {title} {Analytic solutions of
  partial differential equations},\ }\href@noop {} {\bibfield  {journal}
  {\bibinfo  {journal} {University of Leeds, Leeds}\ } (\bibinfo {year}
  {2004})}\BibitemShut {NoStop}%
\bibitem [{\citenamefont {Weinberg}(1972)}]{weinberg1972gravitation}%
  \BibitemOpen
  \bibfield  {author} {\bibinfo {author} {\bibfnamefont {S.}~\bibnamefont
  {Weinberg}},\ }\href@noop {} {\emph {\bibinfo {title} {Gravitation and
  cosmology: principles and applications of the general theory of
  relativity}}}\ (\bibinfo  {publisher} {Wiley},\ \bibinfo {year}
  {1972})\BibitemShut {NoStop}%
\bibitem [{\citenamefont {{Lord Rayleigh}}(1914)}]{rayleigh}%
  \BibitemOpen
  \bibfield  {author} {\bibinfo {author} {\bibnamefont {{Lord Rayleigh}}},\
  }\bibfield  {title} {\bibinfo {title} {On the theory of long waves and
  bores},\ }\href {http://www.jstor.org/stable/93519} {\bibfield  {journal}
  {\bibinfo  {journal} {Proceedings of the Royal Society of London. Series A,
  Containing Papers of a Mathematical and Physical Character}\ }\textbf
  {\bibinfo {volume} {90}},\ \bibinfo {pages} {324} (\bibinfo {year}
  {1914})}\BibitemShut {NoStop}%
\bibitem [{\citenamefont {{Volovik}}(2005)}]{vHe}%
  \BibitemOpen
  \bibfield  {author} {\bibinfo {author} {\bibfnamefont {G.~E.}\ \bibnamefont
  {{Volovik}}},\ }\bibfield  {title} {\bibinfo {title} {{Hydraulic jump as a
  white hole}},\ }\href {https://doi.org/10.1134/1.2166908} {\bibfield
  {journal} {\bibinfo  {journal} {Soviet Journal of Experimental and
  Theoretical Physics Letters}\ }\textbf {\bibinfo {volume} {82}},\ \bibinfo
  {pages} {624} (\bibinfo {year} {2005})}\BibitemShut {NoStop}%
\bibitem [{\citenamefont {Jannes}\ \emph {et~al.}(2011)\citenamefont {Jannes},
  \citenamefont {Piquet}, \citenamefont {Ma\"{\i}ssa}, \citenamefont {Mathis},\
  and\ \citenamefont {Rousseaux}}]{gr1}%
  \BibitemOpen
  \bibfield  {author} {\bibinfo {author} {\bibfnamefont {G.}~\bibnamefont
  {Jannes}}, \bibinfo {author} {\bibfnamefont {R.}~\bibnamefont {Piquet}},
  \bibinfo {author} {\bibfnamefont {P.}~\bibnamefont {Ma\"{\i}ssa}}, \bibinfo
  {author} {\bibfnamefont {C.}~\bibnamefont {Mathis}},\ and\ \bibinfo {author}
  {\bibfnamefont {G.}~\bibnamefont {Rousseaux}},\ }\bibfield  {title} {\bibinfo
  {title} {Experimental demonstration of the supersonic-subsonic bifurcation in
  the circular jump: A hydrodynamic white hole},\ }\href
  {https://doi.org/10.1103/PhysRevE.83.056312} {\bibfield  {journal} {\bibinfo
  {journal} {Phys. Rev. E}\ }\textbf {\bibinfo {volume} {83}},\ \bibinfo
  {pages} {056312} (\bibinfo {year} {2011})}\BibitemShut {NoStop}%
\bibitem [{\citenamefont {Rolley}\ \emph {et~al.}(2006)\citenamefont {Rolley},
  \citenamefont {Guthmann}, \citenamefont {Pettersen},\ and\ \citenamefont
  {Chevallier}}]{He}%
  \BibitemOpen
  \bibfield  {author} {\bibinfo {author} {\bibfnamefont {{\'E}.}~\bibnamefont
  {Rolley}}, \bibinfo {author} {\bibfnamefont {C.}~\bibnamefont {Guthmann}},
  \bibinfo {author} {\bibfnamefont {M.~S.}\ \bibnamefont {Pettersen}},\ and\
  \bibinfo {author} {\bibfnamefont {C.}~\bibnamefont {Chevallier}},\ }\bibfield
   {title} {\bibinfo {title} {The hydraulic jump in liquid helium},\ }\href
  {https://doi.org/10.1063/1.2354642} {\bibfield  {journal} {\bibinfo
  {journal} {AIP Conference Proceedings}\ }\textbf {\bibinfo {volume} {850}},\
  \bibinfo {pages} {141} (\bibinfo {year} {2006})}\BibitemShut {NoStop}%
\bibitem [{\citenamefont {Rousseaux}\ and\ \citenamefont {Kellay}(2020)}]{gr2}%
  \BibitemOpen
  \bibfield  {author} {\bibinfo {author} {\bibfnamefont {G.}~\bibnamefont
  {Rousseaux}}\ and\ \bibinfo {author} {\bibfnamefont {H.}~\bibnamefont
  {Kellay}},\ }\bibfield  {title} {\bibinfo {title} {Classical hydrodynamics
  for analogue space–times: open channel flows and thin films},\ }\href
  {https://doi.org/10.1098/rsta.2019.0233} {\bibfield  {journal} {\bibinfo
  {journal} {Philosophical Transactions of the Royal Society A: Mathematical,
  Physical and Engineering Sciences}\ }\textbf {\bibinfo {volume} {378}},\
  \bibinfo {pages} {20190233} (\bibinfo {year} {2020})}\BibitemShut {NoStop}%
\bibitem [{\citenamefont {Euv\'e}\ \emph
  {et~al.}(2020{\natexlab{b}})\citenamefont {Euv\'e}, \citenamefont
  {Robertson}, \citenamefont {James}, \citenamefont {Fabbri},\ and\
  \citenamefont {Rousseaux}}]{PhysRevLett.124.141101}%
  \BibitemOpen
  \bibfield  {author} {\bibinfo {author} {\bibfnamefont {L.-P.}\ \bibnamefont
  {Euv\'e}}, \bibinfo {author} {\bibfnamefont {S.}~\bibnamefont {Robertson}},
  \bibinfo {author} {\bibfnamefont {N.}~\bibnamefont {James}}, \bibinfo
  {author} {\bibfnamefont {A.}~\bibnamefont {Fabbri}},\ and\ \bibinfo {author}
  {\bibfnamefont {G.}~\bibnamefont {Rousseaux}},\ }\bibfield  {title} {\bibinfo
  {title} {Scattering of co-current surface waves on an analogue black hole},\
  }\href {https://doi.org/10.1103/PhysRevLett.124.141101} {\bibfield  {journal}
  {\bibinfo  {journal} {Phys. Rev. Lett.}\ }\textbf {\bibinfo {volume} {124}},\
  \bibinfo {pages} {141101} (\bibinfo {year} {2020}{\natexlab{b}})}\BibitemShut
  {NoStop}%
\bibitem [{\citenamefont {Montes}\ and\ \citenamefont {Chanson}(1998)}]{chr}%
  \BibitemOpen
  \bibfield  {author} {\bibinfo {author} {\bibfnamefont {J.~S.}\ \bibnamefont
  {Montes}}\ and\ \bibinfo {author} {\bibfnamefont {H.}~\bibnamefont
  {Chanson}},\ }\bibfield  {title} {\bibinfo {title} {Characteristics of
  undular hydraulic jumps: Experiments and analysis},\ }\href
  {https://doi.org/10.1061/(ASCE)0733-9429(1998)124:2(192)} {\bibfield
  {journal} {\bibinfo  {journal} {Journal of Hydraulic Engineering}\ }\textbf
  {\bibinfo {volume} {124}},\ \bibinfo {pages} {192} (\bibinfo {year}
  {1998})}\BibitemShut {NoStop}%
\bibitem [{\citenamefont {Fourdrinoy}\ \emph {et~al.}(2022)\citenamefont
  {Fourdrinoy}, \citenamefont {Robertson}, \citenamefont {James}, \citenamefont
  {Fabbri},\ and\ \citenamefont {Rousseaux}}]{gr3}%
  \BibitemOpen
  \bibfield  {author} {\bibinfo {author} {\bibfnamefont {J.}~\bibnamefont
  {Fourdrinoy}}, \bibinfo {author} {\bibfnamefont {S.}~\bibnamefont
  {Robertson}}, \bibinfo {author} {\bibfnamefont {N.}~\bibnamefont {James}},
  \bibinfo {author} {\bibfnamefont {A.}~\bibnamefont {Fabbri}},\ and\ \bibinfo
  {author} {\bibfnamefont {G.}~\bibnamefont {Rousseaux}},\ }\bibfield  {title}
  {\bibinfo {title} {Correlations on weakly time-dependent transcritical
  white-hole flows},\ }\href {https://doi.org/10.1103/PhysRevD.105.085022}
  {\bibfield  {journal} {\bibinfo  {journal} {Phys. Rev. D}\ }\textbf {\bibinfo
  {volume} {105}},\ \bibinfo {pages} {085022} (\bibinfo {year}
  {2022})}\BibitemShut {NoStop}%
\bibitem [{\citenamefont {Johnson}(1972)}]{viscousjump}%
  \BibitemOpen
  \bibfield  {author} {\bibinfo {author} {\bibfnamefont {R.~S.}\ \bibnamefont
  {Johnson}},\ }\bibfield  {title} {\bibinfo {title} {Shallow water waves on a
  viscous fluid—the undular bore},\ }\href
  {https://doi.org/10.1063/1.1693764} {\bibfield  {journal} {\bibinfo
  {journal} {The Physics of Fluids}\ }\textbf {\bibinfo {volume} {15}},\
  \bibinfo {pages} {1693} (\bibinfo {year} {1972})}\BibitemShut {NoStop}%
\bibitem [{\citenamefont {Steinrück}\ \emph {et~al.}(2003)\citenamefont
  {Steinrück}, \citenamefont {Schneider},\ and\ \citenamefont
  {Grillhofer}}]{Steinruck_2003}%
  \BibitemOpen
  \bibfield  {author} {\bibinfo {author} {\bibfnamefont {H.}~\bibnamefont
  {Steinrück}}, \bibinfo {author} {\bibfnamefont {W.}~\bibnamefont
  {Schneider}},\ and\ \bibinfo {author} {\bibfnamefont {W.}~\bibnamefont
  {Grillhofer}},\ }\bibfield  {title} {\bibinfo {title} {A multiple scales
  analysis of the undular hydraulic jump in turbulent open channel flow},\
  }\href {https://doi.org/10.1016/S0169-5983(03)00041-8} {\bibfield  {journal}
  {\bibinfo  {journal} {Fluid Dynamics Research}\ }\textbf {\bibinfo {volume}
  {33}},\ \bibinfo {pages} {41} (\bibinfo {year} {2003})}\BibitemShut {NoStop}%
\bibitem [{\citenamefont {Kamchatnov}(2000)}]{kam}%
  \BibitemOpen
  \bibfield  {author} {\bibinfo {author} {\bibfnamefont {A.~M.}\ \bibnamefont
  {Kamchatnov}},\ }\href {https://doi.org/10.1142/4513} {\emph {\bibinfo
  {title} {Nonlinear Periodic Waves and Their Modulations}}}\ (\bibinfo
  {publisher} {WORLD SCIENTIFIC},\ \bibinfo {year} {2000})\BibitemShut
  {NoStop}%
\bibitem [{\citenamefont {{Gurevich}}\ and\ \citenamefont
  {{Pitaevski\v\i}}(1974)}]{gurevich}%
  \BibitemOpen
  \bibfield  {author} {\bibinfo {author} {\bibfnamefont {A.~V.}\ \bibnamefont
  {{Gurevich}}}\ and\ \bibinfo {author} {\bibfnamefont {L.~P.}\ \bibnamefont
  {{Pitaevski\v\i}}},\ }\bibfield  {title} {\bibinfo {title} {{Nonstationary
  structure of a collisionless shock wave}},\ }\href@noop {} {\bibfield
  {journal} {\bibinfo  {journal} {Soviet Journal of Experimental and
  Theoretical Physics}\ }\textbf {\bibinfo {volume} {38}},\ \bibinfo {pages}
  {291} (\bibinfo {year} {1974})}\BibitemShut {NoStop}%
\bibitem [{\citenamefont {Arfken}\ \emph {et~al.}(2013)\citenamefont {Arfken},
  \citenamefont {Weber},\ and\ \citenamefont
  {Harris}}]{arfken2013mathematical}%
  \BibitemOpen
  \bibfield  {author} {\bibinfo {author} {\bibfnamefont {G.}~\bibnamefont
  {Arfken}}, \bibinfo {author} {\bibfnamefont {H.}~\bibnamefont {Weber}},\ and\
  \bibinfo {author} {\bibfnamefont {F.}~\bibnamefont {Harris}},\ }\href
  {https://books.google.co.kr/books?id=qLFo\_Z-PoGIC} {\emph {\bibinfo {title}
  {Mathematical Methods for Physicists: A Comprehensive Guide}}}\ (\bibinfo
  {publisher} {Elsevier Science},\ \bibinfo {year} {2013})\BibitemShut
  {NoStop}%
\end{thebibliography}%
\end{document}